\documentclass[
twocolumn,
unsortedaddress,
 amsmath,amssymb,
 aps, physrev,
]{revtex4-2}

\usepackage{graphicx}% Include figure files
\usepackage[english]{babel}
\usepackage{dcolumn}% Align table columns on decimal point
\usepackage{amsmath}
\usepackage{bm}% bold math
\newcommand{\bv}{\bm{v}}

\usepackage{hyperref}% add hypertext capabilities
\usepackage{xcolor}
\newcommand{\bT}{\bm{T}}

\newcommand{\ii}{\mathrm{i}}
\newcommand{\bn}{\hat{\boldsymbol{n}}}
\newcommand{\bt}{\hat{\boldsymbol{t}}}
\newcommand{\btone}{\hat{\boldsymbol{t}}_1}
\newcommand{\bttwo}{\hat{\boldsymbol{t}}_2}
\newcommand{\bU}{\boldsymbol{U}}
\newcommand{\bB}{\boldsymbol{B}}
\newcommand{\bk}{\boldsymbol{k}}
\newcommand{\bE}{\boldsymbol{E}}
\newcommand{\bI}{\mathbf{I}}
\newcommand{\bF}{\boldsymbol{F}}

\newcommand{\avg}[1]{\left\langle #1 \right\rangle}
\newcommand{\jump}[1]{\left[\!\left[#1\right]\!\right]}
\newcommand{\Dt}{\Delta_t}
\newcommand{\bA}{\boldsymbol{A}}

\begin{document}

% \preprint{APS/123-QED}

\title{Turbulence-Driven Corrugation of Collisionless Fast-Magnetosonic Shocks} 

\author{Immanuel Christopher Jebaraj}
\email{immanuel.c.jebaraj@gmail.com}
 \affiliation{Department of Physics and Astronomy, University of Turku, FI-20014 Turun yliopisto, Finland}%Lines break automatically or can be forced with \\

\author{Mikhail Malkov}%
\affiliation{Department of Astronomy and Astrophysics, University of California, San Diego, La Jolla, CA 92093, USA}
\altaffiliation{Eureka Scientific, Inc.,  Oakland, CA 94602, USA}

\author{Nicolas Wijsen}
\affiliation{Center for mathematical Plasma Astrophysics, KU Leuven, Kortrijk/Leuven, Belgium}

\author{Jens Pomoell}
\affiliation{University of Helsinki, 00014 Helsinki, Finland}

\author{Vladimir Krasnoselskikh}
\affiliation{LPC2E/CNRS, UMR 7328, 45071 Orléans, France}
\altaffiliation{Space Sciences Laboratory, University of California, Berkeley, CA 94720, USA}

\author{Nina Dresing}
 \affiliation{Department of Physics and Astronomy, University of Turku, FI-20014 Turun yliopisto, Finland}

\author{Rami Vainio}
 \affiliation{Department of Physics and Astronomy, University of Turku, FI-20014 Turun yliopisto, Finland}

\date{\today}% It is always \today, today,

\begin{abstract}
Collisionless fast-magnetosonic shocks are often treated as smooth, planar boundaries, yet observations point to organized corrugation of the shock surface. A plausible driver is upstream turbulence. Broadband fluctuations arriving at the front can continually wrinkle it, changing the local shock geometry and, in turn, conditions for particle injection and radiation. We develop a linear-Magnetohydrodynamics formulation that treats the shock as a moving interface rather than a fixed boundary. In this approach the shock response can be summarized by an effective impedance determined by the Rankine--Hugoniot base state and the shock geometry, while the upstream turbulence enters only through its statistics. This provides a practical mapping from an assumed incident spectrum to the corrugation amplitude, its drift along the surface, and a coherence scale set by weak damping or leakage. The response is largest when the transmitted downstream fast mode propagates nearly parallel to the shock in the shock frame, which produces a Lorentzian-type enhancement controlled by the downstream normal group speed. We examine how compression, plasma $\beta$, and obliquity affect these corrugation properties and discuss implications for fine structure in heliospheric and supernova-remnant shock emission.
\end{abstract}

%

\maketitle

%\tableofcontents

\section{Introduction}

Collisionless shocks convert directed flow energy into heat, magnetic field compression, waves, suprathermal particles, and radiation \cite{Sagdeev66,Galeev76,Kennel85,Krasnoselskikh13}. The partition of this energy remains unresolved and limits the degree to which we understand fluid and kinetic responses \cite[][]{Sagdeev61,wilson2019,Schwartz2022,Agapitov23,Gedalin23,Gedalin2025}. In the heliosphere, shocks form upstream of planetary magnetospheres or are driven by transients such as coronal mass ejections and corotating interaction regions; they also form in the low corona \cite{Tsurutani1985,Pomoell08,Warmuth15,Kouloumvakos25}. Global behavior is primarily governed by the fast Mach number \(M_f=U_{n1}/c_{f1}\) (ratio of the upstream normal flow speed to the upstream fast magnetosonic speed), the plasma beta \(\beta=2p/B^2\) (thermal-to-magnetic pressure ratio), and the obliquity \(\theta_{Bn}\) (angle between the upstream magnetic field and the shock normal), which separate quasi-parallel and quasi-perpendicular regimes \cite{Kennel88}.

In ideal magnetohydrodynamics (MHD), fast-magnetosonic shocks are infinitesimally thin discontinuities that satisfy the Rankine--Hugoniot (RH) conditions and have no intrinsic surface dynamics \cite{HoffTeller50,Kulikovskii65,Akhiezer75}. Beyond ideal MHD, the front has ion-scale width and dispersion competes with nonlinear steepening, producing whistler precursors and ramp oscillations \cite{Galeev1963,Krasnoselskikh85b,manheimer1985longitudinal,Gedalin98,Krasnoselskikh2002}. Such fine structure is widely observed across heliospheric shocks over a broad range of \(M_f\) and \(\theta_{Bn}\) \cite{Wilson17,Wilson25,Balikhin23,Jebaraj24a}. At high \(M_f\), quasi-perpendicular shocks exhibit ion-scale tangential ripples that propagate along the face \cite{winske1988,Burgess07,Ofman13,Johlander16,Johlander18,Gedalin2024,Gedalin2025} and at even higher \(M_f\) become highly non-stationary leading them to self-reform \cite{Krasnoselskikh85b,Krasnoselskikh2002,lobzin2007}.

Observations indicate that mesoscale nonplanarity is also common. At 1~AU, multi-spacecraft studies report departures from planarity even in weak, quasi-perpendicular events \cite{Neugebauer05,kajdivc2019first}. At planetary bow shocks, ion-scale ripples are routinely resolved \cite{Johlander16,Johlander18} while larger scale ``breathing'' has also been reported \cite{Huterer97,Horbury02,cheng2025bow}. In the low corona, type~II radio emission often shows multiple, closely spaced lanes and moving sub-sources that drift along the shock surface \cite{Zlobec93,Morosan25}. The inferred spacings exceed ion scales and point to organized corrugation on MHD scales, while global deformations by streamers explain broader trends but not the narrow spacing or rapid reconfiguration \cite{Jebaraj21,Jebaraj23b,Kouloumvakos21,Wijsen23,Wijsen25,Magdalenic20}.

The upstream solar wind is turbulent from the corona outward \cite{Bruno13}, with dominantly Alfv\'enic, weakly compressive fluctuations near the Sun and increasing compressive power and intermittency with distance \cite{Chen20,Sioulas22}. Transmission, refraction, and mode conversion of this broadband turbulent plasma at shocks have been extensively studied using observations, theory, and simulations \cite{Kennel1982,BoillatRuggeri1979,Achterberg86,Vainio98,Vainio99,Zank02,laming2015wave,Zank2021,Gedalin23,Giacalone05,Demidem2018,Naka22,Sishtla23,Trotta2023,Turc23}. Most existing treatments, however, regard the shock as a fixed interface and focus on the evolution of downstream waves. The surface response to realistic upstream fluctuations remains under explored particularly knowing that it plays a leading order role in this boundary value problem \cite{lubchich2005magnetohydrodynamic,fraschetti2013turbulent,lemoine2016corrugation,Zank2021}. Limited observations point to mesoscale corrugations with correlation lengths near \(10^6\)~km at 1~AU \cite{Bale99,Neugebauer05,Pulupa10} and simulations suggest that incident turbulence can drive such corrugations above ion scales \cite{Giacalone05,Demidem2018,Guo21,Trotta2023}.

Shocks can be major particle accelerators, yet the role of nonplanarity in that process remains poorly constrained. Most theoretical work since the advent of diffusive shock acceleration (DSA) has focused on planar shocks \cite{Krymskii77,Axford77,Blandford78,Bell78,Malkov01},  leaving the effects of local curvature and surface variability comparatively unexplored. Observations, however, suggest that small, transient variations in local obliquity and compression modulate particle acceleration \cite{Sandroos06,Guo10,Xu25,Trotta25} and produce localized radiation hot spots \cite{Bale99,Kuncic02,Pulupa10,Cairns25}, consistent with spatially structured sources in heliospheric shocks and the recurring multi-lane morphology of type~II bursts \cite{Jebaraj23,Jebaraj24b,Wilson25,Magdalenic20,Normo25,Morosan25,zucca2025source}.

Beyond heliospheric shocks, comparable behavior is reported at astrophysical shocks \cite{Bamba03,Uchiyama07}. In young supernova remnants (SNRs) such as Tycho, time–variable X–ray stripes exhibit characteristic spacings of order \(10^{10}\,\mathrm{km}\) \cite{eriksen2011evidence}. As in coronal shocks, the stripes evolve on observation timescales (\(\sim 14\) measurements over 15 years), indicating a dynamic modulation of conditions at the shock surface \cite{okuno2020time}. This phenomenology is consistent with shocks propagating through a turbulent medium that imprint coherent corrugation of the front.

Here, we examine how upstream fluctuations perturb the shock interface. We investigate whether such perturbations can drive coherent, long-lived corrugations and how their characteristics depend on the compression, \(\beta\), and \(\theta_{Bn}\). We build an interfacial framework in linear MHD, resolve the transmitted fast-mode kinematics, and examine implications for the amplitude of corrugations, their speed, and the coherence. In this formulation of the classic boundary value problem, we isolate the shock interface as a dynamic coupling layer between upstream turbulence and mesoscale corrugation.

The remainder of the paper is organized as follows. In Sec.~\ref{sec:formulation} we formulate shock corrugation as a driven moving-boundary problem, derive the linear interfacial response in Fourier space, and reduce the boundary system to an effective impedance relation. In Sec.~\ref{sec:discussion} we evaluate the resulting corrugation characteristics for representative fast shocks driven by broadband upstream fluctuations. In Sec.~\ref{sec:accel_implications} we discuss the implications of a perturbed shock boundary for particle dynamics. Finally, in Sec.~\ref{sec:conclusions} we summarize the main results, discuss limitations of the idealized closure, and outline observational and numerical tests.

\section{Formulation of the problem}\label{sec:formulation}

% \begin{figure}
% \centering
% \includegraphics[width=0.48\columnwidth]{Figures/Res_cone.eps}%
% \hfill
% \includegraphics[width=0.48\columnwidth]{Figures/Res_cone2.eps}
% \caption{(left panel) Schematic of the interface formulation in the shock rest frame. A turbulent upstream flow convects broadband fluctuations toward an unperturbed planar front ($\hat{\boldsymbol n}$) which becomes corrugated. The local shock normal is $\hat{\boldsymbol n}-\nabla_{\!\perp}\zeta$.
% (b) Resonance–cone geometry in $(k_{n2},\boldsymbol{k}_\perp)$ for the
% transmitted downstream fast mode, showing the locus where
% $v_{g,n2}=0$ and the surface response peaks.}
% \label{fig:rescone}
% \end{figure}

%========================
% Figure 0a — Schematic of the problem
%========================
\begin{figure}
\centering
\includegraphics[width=0.9\columnwidth]{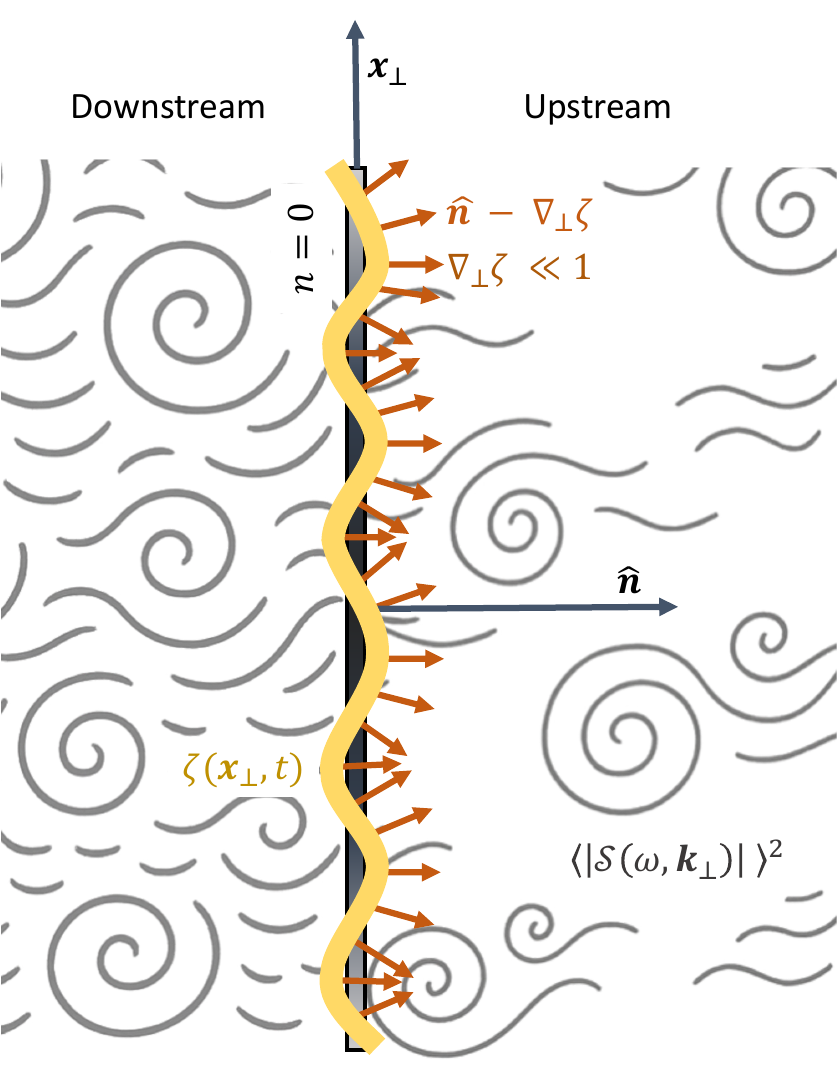}
\caption{Schematic of the interface formulation in the shock rest frame. A turbulent upstream flow convects broadband fluctuations toward an initially planar shock with normal $\hat{\boldsymbol n}$, which is displaced into a corrugated surface described by $\zeta(\boldsymbol{x}_\perp,t)$. The local shock normal is then $\hat{\boldsymbol n}-\nabla_{\!\perp}\zeta$.}
\label{fig:rescone}
\end{figure}

\subsection{Moving-interface linearization and admissible downstream response}

We model the coupling between upstream turbulence and a shock front with linear MHD in the small–amplitude, small–slope limit, working in the shock rest frame. Figure~\ref{fig:rescone} provides a schematic of this setup. We treat the shock as a moving, planar interface in a magnetized plasma. An incident disturbance with frequency $\omega$ and wavevector $\bk$ meets this interface. Rather than holding the interface fixed, we take the surface position $\zeta(\boldsymbol{x}_\perp,t)$ as a dynamical variable. Writing the conservation laws in flux form and applying the Hadamard jump conditions \cite{Hadamard1903} to the moving level set yields a closed linear evolution for $\zeta$ driven by the incident field. The unperturbed shock satisfies the RH conditions; linearization about this base state separates the susceptibility of the surface from the statistics of the upstream fluctuations, and the physically relevant downstream responses are selected to propagate or decay away from the interface (in the classical Lopatinskii-Shapiro sense \cite{lopatinskii1953,Shapiro1953}). 

The result is a scalar boundary evolution of the form

\[
\mathcal{Z}(\omega,\bk_\perp)\,\zeta(\omega,\bk_\perp)=\mathcal{S}(\omega,\bk_\perp).
\]

Here, the interfacial impedance \(\mathcal{Z}(\omega,\bk_\perp)\)  depends only on the base state and geometry;  the scalar drive \(\mathcal{S}(\omega,\bk_\perp)\) is obtained by projecting the upstream perturbations through the same boundary operators; the radiation condition (no incoming energy from downstream) and the grazing set \(v_{g,n2}=0\) are posed in the shock frame \cite{kontorovich1959interaction,BoillatRuggeri1979,AsseoBerthomieu1970,lubchich2005magnetohydrodynamic,Zank2021}.

We adopt ideal MHD with \(\mu_0=1\) (Heaviside–Lorentz units), so \(v_A=|\bB|/\sqrt{\rho}\) \cite{LandauFM59}. Coordinates are adapted to a steady planar shock at \(n=\bn\!\cdot\!\boldsymbol{x}=0\), with unit normal \(\bn\) and orthonormal tangents \(\{\btone,\bttwo\}\) (as in Figure~\ref{fig:rescone}). Any vector \(\bA\) is decomposed as \(A_n=\bA\!\cdot\!\bn\) and \(\bA_\perp=\bA-A_n\bn\). Upstream (downstream) quantities carry subscript \(1\) (\(2\)); the subscript \(0\) denotes the unperturbed RH base state. The obliquity is \(\cos\theta_{Bn}=\hat{\bB}\!\cdot\!\bn\). For later use we define the along–surface wavevector \(\bk_\perp=\bk-(\bk\!\cdot\!\bn)\bn\). 
Tangential phase matching sets refraction, flux continuity sets reflection and transmission, and the radiation condition selects admissible downstream modes. Within this framework, all geometry and RH dependence reside in \(\mathcal{Z}\), while the chosen upstream spectra enter only through \(\mathcal{S}\). The conservation laws read

\begin{equation}\label{eq:mass}
\partial_t \rho + \nabla\!\cdot(\rho\,\bU) = 0,
\end{equation}

\begin{equation}\label{eq:momentum}
\partial_t(\rho \bU) + \nabla\!\cdot\!\Big[\rho\,\bU\bU + \Big(p+\tfrac12 B^2\Big)\bI - \bB\bB\Big] = 0,
\end{equation}

\begin{equation}\label{eq:induction}
\partial_t \bB - \nabla\times(\bU\times\bB) = 0,\qquad \nabla\!\cdot\bB=0,
\end{equation}

\begin{equation}\label{eq:energy}
\begin{aligned}
\partial_t\Big(\tfrac12\rho U^2 + \tfrac{p}{\gamma-1} + \tfrac12 B^2\Big)
&+ \nabla\!\cdot\!\Big[
   \Big(\tfrac12\rho U^2 + \tfrac{\gamma}{\gamma-1}p + B^2\Big)\bU \\
&\qquad\qquad - (\bU\!\cdot\!\bB)\,\bB
\Big] = 0 .
\end{aligned}
\end{equation}

Using

\begin{equation}\label{eq:curl_to_div}
\nabla\times(\bU\times\bB)
= -\,\nabla\!\cdot(\bU\bB-\bB\bU),
\end{equation}

Eq.~\eqref{eq:induction} can be written in flux form such that all governing equations are divergences of fluxes.

\begin{equation}\label{eq:induction_flux}
\partial_t \bB + \nabla\!\cdot(\bU\bB-\bB\bU)=0 .
\end{equation}

Integrating Eqs.~\eqref{eq:mass}--\eqref{eq:energy} over a thin pillbox straddling \(n=0\) and shrinking its thickness yields continuity of normal fluxes (the RH base state). From Eq.~\eqref{eq:mass} and \(\nabla\!\cdot\bB=0\),

\begin{equation}\label{eq:RH_mass}
\jump{\rho U_n}=0,\qquad \jump{B_n}=0.
\end{equation}

With \(\bT=\rho\,\bU\bU + (p+\tfrac12 B^2)\bI - \bB\bB\), the tangential and normal momentum conditions are

\begin{equation}\label{eq:RH_tan_mom}
\jump{\rho U_n \bU_\perp - B_n \bB_\perp}=\boldsymbol{0},
\end{equation}

\begin{equation}\label{eq:RH_norm_mom}
\jump{\rho U_n^2 + p + \tfrac12 B^2 - B_n^2}=0.
\end{equation}

From Eq.~\eqref{eq:curl_to_div}, the tangential induction condition is

\begin{equation}\label{eq:RH_Etan}
\jump{U_n \bB_\perp - \bU_\perp B_n}=\boldsymbol{0}.
\end{equation}

The base state is parameterized by \(r=\rho_2/\rho_1\), implying \(U_{n2}=U_{n1}/r\) and \(B_{n2}=B_{n1}\). Solving Eq.~\eqref{eq:RH_tan_mom} and Eq.~\eqref{eq:RH_Etan} for \((\bU_{2\perp},\bB_{2\perp})\) along any fixed \(\bt\) yields the \(2\times 2\) system

\begin{equation}\label{eq:Mt}
\begin{bmatrix}\rho_2 U_{n2} & -B_n\\ -B_n & U_{n2}\end{bmatrix}
\begin{bmatrix}\bt\!\cdot\!\bU_{2\perp}\\ \bt\!\cdot\!\bB_{2\perp}\end{bmatrix}
=
\begin{bmatrix}\bt\!\cdot\!\bT_1\\ \bt\!\cdot\!\bE_1\end{bmatrix},
\end{equation}

where

\begin{equation}\label{eq:invariants}
\begin{aligned}
&\bT_1=\rho_1 U_{n1}\bU_{1\perp}-B_n \bB_{1\perp},\\
&\bE_1=U_{n1}\bB_{1\perp}-\bU_{1\perp}B_n.  
\end{aligned}
\end{equation}

The determinant is

\begin{equation}\label{eq:detMt}
\begin{aligned}
\det
&= \rho_2\!\left(U_{n2}^2 - v_{A,n2}^2\right) \\
& = \rho_2 U_{n2}^2 - B_n^2 \\
& = \frac{\rho_1 U_{n1}^2}{r} - B_n^2 \equiv -\Delta_t, \\
\\ \Delta_t
&= \rho_2\!\left(v_{A,n2}^2 - U_{n2}^2\right)
 = B_n^2 - \frac{\rho_1 U_{n1}^2}{r}.
\end{aligned}
\end{equation}

where, \(v_{A,n2}^2 \equiv \frac{B_n^2}{\rho_2}\). Inverting gives the explicit transmissions

\begin{equation}\label{eq:B2perp}
\bB_{2\perp}=\frac{B_n^2-\rho_1 U_{n1}^2}{\Dt}\,\bB_{1\perp},
\end{equation}
\begin{equation}\label{eq:U2perp}
\bU_{2\perp}=\bU_{1\perp}+\frac{U_{n1}B_n\big(\tfrac{1}{r}-1\big)}{\Dt}\,\bB_{1\perp}.
\end{equation}

The normal momentum balance then yields

\begin{equation}\label{eq:p2}
p_2 = p_1
+ \frac{(\rho_1 U_{n1})^2}{\rho_1}
- \frac{(\rho_2 U_{n2})^2}{\rho_2}
+ \tfrac12\Big(|\bB_{1\perp}|^2-|\bB_{2\perp}|^2\Big),
\end{equation}

and the total energy flux

\begin{equation}\label{eq:FEdef}
\bF_E
=\Big(\tfrac12\rho U^2+\tfrac{\gamma}{\gamma-1}p+B^2\Big)\bU
-(\bU\!\cdot\!\bB)\,\bB
\end{equation}

must satisfy continuity of its normal component. We collect this RH  closure into

\begin{equation}\label{eq:F_lam}
\mathcal{F}_{\mathrm{lam}}
\big(r;\ \rho_1,p_1,\bU_1,\bB_1,\gamma\big)
\equiv
\big(\bF_E\!\cdot\!\bn\big)_2(r)
-
\big(\bF_E\!\cdot\!\bn\big)_1
=0,
\end{equation}

where $(\cdot)_1$ denotes upstream quantities and $(\cdot)_2(r)$ are the downstream quantities expressed in terms of $r=\rho_2/\rho_1$ via Eqs.~\eqref{eq:RH_mass}--\eqref{eq:p2}. Solving $\mathcal{F}_{\mathrm{lam}}=0$ fixes $r$ for given upstream state and $\gamma$.

To endow the interface with dynamics we represent the moving surface by the level set

\begin{equation}\label{eq:Phi_def}
\Phi(\boldsymbol{x},t)
= n(\boldsymbol{x}) - \zeta(\boldsymbol{x}_\perp,t)
= \bn\!\cdot\!\boldsymbol{x} - \zeta(\boldsymbol{x}_\perp,t),
\end{equation}

where $n(\boldsymbol{x})=\bn\!\cdot\!\boldsymbol{x}$ is the coordinate along the fixed unit normal $\bn$ of the unperturbed shock (so $n=0$ is the base plane) and $\boldsymbol{x}_\perp$ are the in-plane coordinates along $\{\btone,\bttwo\}$. The perturbed interface is $\Phi=0$, i.e. $n=\zeta(\boldsymbol{x}_\perp,t)$.

For small slopes $|\nabla_{\!\perp}\zeta|\ll1$,

\begin{equation}
\nabla\Phi = \bn - \nabla_{\!\perp}\zeta
+ \mathcal{O}(|\nabla_{\!\perp}\zeta|^2),
\quad
-\partial_t\Phi = \partial_t\zeta,
\end{equation}

so to leading order the interface moves with normal speed $\partial_t\zeta$ and has local normal $\bn-\nabla_{\!\perp}\zeta$.

For any conservation law \(\partial_t Q+\nabla\!\cdot\bF=0\), integrating across the moving surface gives Hadamard’s condition \cite{Hadamard1903,TruesdellToupin60}

\begin{equation}\label{eq:Hadamard}
\jump{\bF\!\cdot\nabla\Phi}+\jump{Q}\,\partial_t\Phi=0.
\end{equation}

Linearizing about the base state where \(\jump{\bF_0\!\cdot\bn}=0\) yields

\begin{equation}\label{eq:Hadamard_lin}
\jump{\delta(\bF\!\cdot\bn)}
-(\partial_t\zeta)\,\jump{Q_0}
-\big(\nabla_{\!\perp}\zeta\big)\!\cdot\!\jump{\bF_{0\perp}}
=0.
\end{equation}

The first term in Eq.~\eqref{eq:Hadamard_lin} is the mismatch of perturbed normal fluxes. The $(\partial_t\zeta)\jump{Q_0}$ term is advection of the base jump by normal motion of the interface. The $(\nabla_{\!\perp}\zeta)\!\cdot\!\jump{\bF_{0\perp}}$ term arises from tilt: rotating the normal by $-\nabla_{\!\perp}\zeta$ mixes any base tangential flux jump into the effective normal flux.

Applying Eq.~\eqref{eq:Hadamard_lin} to Eqs.~\eqref{eq:mass}--\eqref{eq:induction} produces the exact linear interfacial relations

\begin{equation}\label{eq:BC_mass}
\jump{\delta(\rho U_n)}
-(\partial_t\zeta)\,\jump{\rho}
-\big(\nabla_{\!\perp}\zeta\big)\!\cdot\!\jump{\rho\,\bU_\perp}
=0.
\end{equation}

\begin{equation}\label{eq:BC_Bn}
\jump{\delta B_n}
-\big(\nabla_{\!\perp}\zeta\big)\!\cdot\!\jump{\bB_{0\perp}}
=0.
\end{equation}

Here \(\delta B_n\) denotes the Eulerian bulk perturbation of the normal field component; the second term is the geometric tilt correction from the perturbed normal. We keep this tilt term explicitly, and it enters the \(\zeta\)-column \(\bm m_\zeta\). 

% Here \(\delta B_n\) includes the contribution from the interface tilt, i.e. the \(-(\nabla_{\!\perp}\zeta)\!\cdot\!\jump{\bB_{0\perp}}\) term is absorbed into \(\delta B_n\).

\begin{equation}\label{eq:BC_tan_mom}
\begin{aligned}
&\bt\!\cdot\!\jump{\delta\!\big(\rho U_n \bU_\perp - B_n \bB_\perp\big)}
-(\partial_t\zeta)\,\bt\!\cdot\!\jump{\rho\,\bU_\perp} \\
&\quad
-\big(\nabla_{\!\perp}\zeta\big)\!\cdot\!\jump{\big(\bT_0^{\top}\bt\big)_{\!\perp}}
= 0.
\end{aligned}
\end{equation}

Here, \((\bt=\btone,\bttwo)\).

\begin{equation}\label{eq:BC_norm_mom}
\jump{\delta(\rho U_n^2 + p + \tfrac12 B^2 - B_n^2)}=0,
\end{equation}

\begin{equation}\label{eq:BC_tan_ind}
\begin{aligned}
&\bt\!\cdot\!\jump{\delta\!\big(U_n\bB_\perp-\bU_\perp B_n\big)}
-(\partial_t\zeta)\,\bt\!\cdot\!\jump{\bB_\perp} \\
&\quad
-\big(\nabla_{\!\perp}\zeta\big)\!\cdot\!
\jump{\big((\bU_0\bB_0-\bB_0\bU_0)^{\top}\bt\big)_{\!\perp}}
= 0.
\end{aligned}
\end{equation}

% The explicit variation entering Eq.~\eqref{eq:BC_tan_ind} is

% \begin{equation}\label{eq:var_Etan}
% \begin{aligned}
% \delta\!\left(U_n\bB_\perp - \bU_\perp B_n\right)
% &= (\delta U_n)\,\bB_\perp + U_n\,\delta\bB_\perp \\
% &\quad - (\delta\bU_\perp) B_n - \bU_\perp\,\delta B_n .
% \end{aligned}
% \end{equation}

% and analogous first-order expansions are used elsewhere.

Bulk fluctuations on each side are represented as linear MHD eigenmodes with plane-wave dependence \(\exp\{\ii(\bk\!\cdot\!\boldsymbol{x}-\omega t)\}\) and Doppler-shifted frequency \(\omega'=\omega-\bk\!\cdot\!\bU\). Linearizing Eqs.~\eqref{eq:mass}--\eqref{eq:induction} about a uniform medium gives

\begin{equation}\label{eq:Lmass}
-\,\ii \omega' \,\delta \rho + \ii \rho\, \bk\!\cdot\!\delta\bU = 0,
\end{equation}

\begin{equation}\label{eq:Lmom}
-\,\ii \omega' \,\rho \,\delta\bU = -\,\ii \bk\,\delta p + \ii (\bk\times \delta\bB)\times \bB,
\end{equation}

\begin{equation}\label{eq:Lind}
-\,\ii \omega' \,\delta\bB = \ii \bk\times(\delta\bU\times \bB),\qquad \bk\!\cdot\!\delta\bB=0,
\end{equation}

\begin{equation}\label{eq:Lclosure}
\delta p = c_{\mathrm s}^2\,\delta\rho + p_S\,\delta S,\qquad 
c_{\mathrm s}^2=\left(\frac{\partial p}{\partial\rho}\right)_{S}=\gamma p/\rho,
\end{equation}

Here, \(p_S\equiv(\partial p/\partial S)_\rho\), and

\begin{equation}\label{eq:Lentropy}
-\,\ii\omega'\,\delta S=0.
\end{equation}

For the entropy/contact mode (denoted by $j=\mathrm{ent}$) we take the eigenvector at the interface to satisfy

\begin{equation}\label{eq:entropy_mode_def}
(\delta\bU_2)^{(\mathrm{ent})}=\boldsymbol{0},\qquad
(\delta\bB_2)^{(\mathrm{ent})}=\boldsymbol{0},\qquad
(\delta p_2)^{(\mathrm{ent})}=0,
\end{equation}

with $(\delta\rho_2)^{(\mathrm{ent})}\neq 0$ and $\delta S_2^{(\mathrm{ent})}$ chosen so that \eqref{eq:Lclosure} gives $(\delta p_2)^{(\mathrm{ent})}=0$.

Eliminating \(\delta\rho\), \(\delta p\), and \(\delta\bB\) yields the compressive dispersion polynomial

\begin{equation}\label{eq:quartic}
\omega'^4 - (c_{\mathrm s}^2+v_A^2)k^2\,\omega'^2
  + c_{\mathrm s}^2(\bk\!\cdot\!\boldsymbol{v}_A)^2 k^2 = 0,
\end{equation}

and the Alfv\'en branch

\begin{equation}\label{eq:alfven_branch}
\begin{aligned}
\omega'_A &= \sigma\,\bk\!\cdot\!\boldsymbol{v}_A,\\
\delta\rho &= 0, \qquad \delta p = 0,\\
\delta\bB &= -\,\frac{\bk\!\cdot\!\bB}{\omega'_A}\,\delta\bU 
           \;=\; -\,\sigma\,\sqrt{\rho}\,\delta\bU,\\
\delta\bU &\perp \bk, \qquad \delta\bU \perp \bB .
\end{aligned}
\end{equation}

with \(\sigma = \pm1\). The group velocity follows from the implicit function \(D(\omega',\bk)=0\) with

\begin{equation}\label{eq:D_def}
D(\omega',\bk)=\omega'^4-(c_{\mathrm s}^2+v_A^2)k^2\,\omega'^2+c_{\mathrm s}^2(\bk\!\cdot\!\boldsymbol{v}_A)^2 k^2.
\end{equation}

For later use, the required derivatives are

\begin{equation}\label{eq:D_partials}
\begin{aligned}
\partial_{\omega'} D &= 4\omega'^3 - 2(c_{\mathrm s}^2+v_A^2)\,k^2\,\omega',\\[2pt]
\partial_{\bk} D &= -2(c_{\mathrm s}^2+v_A^2)\,\omega'^2\,\bk
+ 2c_\mathrm{s}^2(\bk\!\cdot\!\boldsymbol{v}_A)\,\boldsymbol{v}_A\,k^2 \\
&\quad + 2c_\mathrm{s}^2(\bk\!\cdot\!\boldsymbol{v}_A)^2\,\bk .
\end{aligned}
\end{equation}

The group velocity and its normal component are thus

\begin{equation}\label{eq:vg}
\boldsymbol{v}_g = \frac{\partial \omega}{\partial \bk}=\bU-\frac{\partial_{\bk} D}{\partial_{\omega'}D},\qquad v_{g,n}=\bn\!\cdot\!\frac{\partial \omega}{\partial \bk}.
\end{equation}

For the compressive branches, writing \(\delta\bU=\alpha\,\hat{\bk}+\eta\,\hat{\bm{\xi}}\) with \(\hat{\bm{\xi}}\) the unit vector in the \((\bk,\bB)\) plane orthogonal to \(\hat{\bk}\), one finds the polarization ratio

\begin{equation}
\label{eq:beta_alpha}
\frac{\eta}{\alpha}
= -\,\frac{\omega'^2 - c_{\mathrm s}^2 k^2 - v_A^2 k_\perp^2}{v_A^2 k_\parallel k_\perp}
= -\,\frac{v_A^2 k_\parallel k_\perp}{\omega'^2 - v_A^2 k_\parallel^2}.
\end{equation}

with \(k_\parallel=\bk\!\cdot\!\hat{\bB}\) and \(k_\perp^2=k^2-k_\parallel^2\). The derivation and algebraic equivalences that lead to Eq.~\eqref{eq:beta_alpha} are given in Appendix~\ref{app:pol}, Eq.~\eqref{app:eq:beta_over_alpha_master}, where the reduction with the dispersion relation is shown explicitly.

At the interface, the upstream fluctuation field is decomposed into an Alfv\'enic part obeying Eq.~\eqref{eq:alfven_branch} and a compressive part on either magnetosonic branch with polarization set by Eq.~\eqref{eq:beta_alpha}. For the Alfv\'en part, defining \(\hat{\bm e}_A=(\bk\times\bB_1)/|\bk\times\bB_1|\), the normal/tangential velocity projections at the interface are

\begin{equation}\label{eq:U1nA}
\delta U_{n1}^{(A)}
= \delta U_1^{(A)}\,\frac{\bn\!\cdot\!(\bk\times\bB_1)}{|\bk\times\bB_1|}
= \delta U_1^{(A)}\,\frac{(\bk_\perp\times\bB_{1\perp})\!\cdot\!\bn}{|\bk\times\bB_1|}.
\end{equation}

\begin{equation}\label{eq:U1tA}
\delta U_{1t}^{(A)}
=\delta U_1^{(A)}\,\bt\!\cdot\!\hat{\bm e}_A
=\delta U_1^{(A)}\,\frac{(\bt\times\bk)\!\cdot\!\bB_1}{|\bk\times\bB_1|}.
\end{equation}

The Alfv\'en polarization $\hat{\bm e}_A=(\bk\times\bB_1)/|\bk\times\bB_1|$ is undefined when $\bk\parallel\bB_1$ because then $\bk\times\bB_1=\boldsymbol{0}$. In that parallel case the Alfv\'en branch still exists: $\delta\bU$ must remain perpendicular to $\bk$ (and hence to $\bB_1$), but its direction within the plane orthogonal to $\bk$ is not unique.

For the compressive part, using Eq.~\eqref{eq:Lmass} and Eq.~\eqref{eq:Lclosure},

\begin{equation}\label{eq:Comp_up_rho_p}
\delta\rho_1^{(C)}=\frac{\rho_1}{\omega_1'}\,\bk\!\cdot\!\delta\bU_1^{(C)}=\frac{\rho_1 k}{\omega_1'}\,\alpha,\qquad \delta p_1^{(C)}= c_{\mathrm s1}^2\,\delta\rho_1^{(C)},
\end{equation}

and from Eq.~\eqref{eq:Lind}

\begin{equation}\label{eq:Comp_up_B}
\delta\bB_1^{(C)}=\frac{1}{\omega_1'}\Big[\bB_1\,(\bk\!\cdot\!\delta\bU_1^{(C)})-\delta\bU_1^{(C)}(\bk\!\cdot\!\bB_1)\Big].
\end{equation}

These relations are used only to form the upstream forcing; the shock face response (interfacial susceptibility) is independent of upstream amplitudes. The explicit quadratic reduction that yields the Alfv\'enic and compressive weights used later, including the \(\cos^2\theta_{Bn}\) and \(\sin^2\theta_{Bn}\) factors, is collected in Appendix~\ref{app:weights}.

On the downstream side we fix $(\omega,\bk_\perp)$ and select admissible eigenmodes in the shock frame. We orient the unit normal $\bn$ from region~1 to region~2, so the downstream half–space is $n>0$. For propagating branches we retain those whose energy flux points away from the interface—equivalently $v_{g,n2}>0$ for positive wave–action density (i.e., $\partial_{\omega'} D_2>0$). For evanescent branches (with no real normal group velocity), we select the root with imaginary normal wavenumber $k_{n2}=i\,q$, $q>0$, so that $e^{i k_{n2} n}=e^{-q n}$ decays into region~2. This is the (Kreiss–)Lopatinskii admissibility rule for hyperbolic initial–boundary problems (see \cite{Kreiss70,majda83}; cf.\ the classical Lopatinskii–Shapiro boundary reduction \cite{lopatinskii1953,Shapiro1953}).\footnote{This outgoing/decaying selection is equivalent to the multidimensional Lopatinskii admissibility condition \cite{majda83,metivier2000block,metivier2001stability}. Its equivalence to a unit–wave–action–flux normalization, well conditioned near grazing, is summarized in Appendix~\ref{app:normalization}.}

Denote the admissible downstream modes by $j$, with normal wavenumbers $k_{n2}^{(j)}$ obtained from the magnetosonic quartic \eqref{eq:quartic}, the Alfv\'en branch \eqref{eq:alfven_branch}, and the entropy/contact branch \eqref{eq:entropy_mode_def}, with corresponding eigenvectors $\psi_2^{(j)}=(\delta\rho_2,\delta\bU_2,\delta\bB_2,\delta p_2)^{(j)}$. The downstream perturbation is then

\[
\psi_2(\omega,\bk_\perp;n)
=\sum_{j=1}^{N} a_j\,\psi_2^{(j)}\,e^{\,i k_{n2}^{(j)} n}, \qquad N=7
\]

where the complex amplitudes $a_j$ absorb any eigenvector normalization. Accordingly, in ideal adiabatic MHD we take $N=7$: fast$_\pm$, slow$_\pm$, Alfv\'en$_\pm$, and the entropy/contact mode.

% On the downstream side, admissible eigenmodes at fixed \((\omega,\bk_\perp)\) are selected by a radiation condition in the shock frame: retain those with energy flux (equivalently $v_{g,n2}$ for positive wave–action density, i.e. $\partial_{\omega'}D_2>0$) pointing away from the interface; exclude incoming roots. For evanescent branches, select the root with imaginary normal wavenumber \(k_{n2}=\ii\kappa\) and $\kappa>0$, so the field decays away from the interface. 
% Denote the admissible modes by \(j\), with normal wavenumbers \(k_{n2}^{(j)}\) from Eq.~\eqref{eq:quartic} and corresponding eigenvectors \(\psi_2^{(j)}=(\delta\rho_2,\delta\bU_2,\delta\bB_2,\delta p_2)^{(j)}\). 
% The downstream perturbation is expressed as a superposition of these eigenmodes,
% \[
% \psi_2(\omega,\bk_\perp;n)
% =\sum_{j=1}^{N} a_j\,\psi_2^{(j)}\,e^{\ii k_{n2}^{(j)} n},
% \]
% where $a_j$ are complex amplitudes and the normalization of each $\psi_2^{(j)}$ is arbitrary—any rescaling can be absorbed into $a_j$. 
% This outgoing/decaying selection is equivalent to the multidimensional Lopatinski admissibility condition \cite[see][]{majda83,metivier2000block,metivier2001stability}.  Equivalence to a unit–wave–action–flux normalization, well conditioned near grazing, is summarized in Appendix\ref{app:normalization}.

To enforce Eqs.~\eqref{eq:BC_mass}–\eqref{eq:BC_tan_ind}, the boundary operators are applied row by row to each eigenmode $\psi_2^{(j)}$, forming the matrix $\mathsf M_{\mathrm{dd}}$. The same operators acting on the kinematic terms $-(\partial_t\zeta)\jump{Q_0}$ and $-(\nabla_{\!\perp}\zeta)\!\cdot\!\jump{\bF_{0\perp}}$, after Fourier substitution $\partial_t\mapsto -i\omega$ and $\nabla_{\!\perp}\mapsto i\bk_\perp$, produce the $\zeta$-column $\bm m_\zeta$. Acting on the upstream fluctuations gives the forcing $(\mathbf f_{\mathrm d},f_\zeta)$. The result is

\begin{equation}\label{eq:blocks}
\mathsf{M}_{\mathrm{dd}}\,\mathbf{a}
+ \bm{m}_\zeta\,\zeta
= \mathbf{f}_{\mathrm{d}},
\qquad
\bm{\ell}_{\mathrm{d}}^{\top}\mathbf{a}
+ \ell_\zeta\,\zeta
= f_\zeta,
\end{equation}

where the second (scalar) equation represents one remaining independent boundary condition not already included in $\mathsf{M}_{\mathrm{dd}}$, with $\bm{\ell}_{\mathrm{d}}^{\top}$ its coefficients on $\mathbf a$, $\ell_\zeta$ its coefficient on $\zeta$, and $f_\zeta$ the corresponding upstream forcing. \footnote{In \eqref{eq:blocks} we take the scalar equation to be the linearized Hadamard condition for the energy conservation law \eqref{eq:energy}, i.e. continuity of the normal component of the energy flux across the moving interface. With $Q_E=\tfrac12\rho U^2+\tfrac{p}{\gamma-1}+\tfrac12 B^2$ and $\bF_E$ given by \eqref{eq:FEdef}, Eq.~\eqref{eq:Hadamard_lin} gives \(\jump{\delta(\bF_E\!\cdot\!\bn)}
-(\partial_t\zeta)\,\jump{Q_{E0}}
-\big(\nabla_{\!\perp}\zeta\big)\!\cdot\!\jump{(\bF_{E0})_{\!\perp}}=0.
\) Upon Fourier substitution $\partial_t\mapsto -i\omega$ and $\nabla_{\!\perp}\mapsto i\bk_\perp$, this provides the scalar row $(\bm\ell_d^\top,\ell_\zeta,f_\zeta)$ used in \eqref{eq:blocks}.}

Eliminating $\mathbf a$ gives

\begin{equation}\label{eq:Schur}
\Big(\ell_\zeta-\bm{\ell}_{\mathrm{d}}^{\top}\mathsf{M}_{\mathrm{dd}}^{-1}\bm{m}_\zeta\Big)\,\zeta
=
f_\zeta-\bm{\ell}_{\mathrm{d}}^{\top}\mathsf{M}_{\mathrm{dd}}^{-1}\mathbf{f}_{\mathrm{d}}.
\end{equation}

\subsection{Interfacial impedance \texorpdfstring{$\mathcal{Z}(\omega,\mathbf k_\perp)$}{Z(omega, k-perp)} and grazing-resonance structure}

We now reduce the boundary system to an effective scalar relation between the shock displacement and the imposed upstream forcing, $\mathcal Z(\omega,\mathbf k_\perp)\,\zeta(\omega,\mathbf k_\perp)= \mathcal S(\omega,\mathbf k_\perp)$, which defines the interfacial impedance $\mathcal Z$ and source term $\mathcal S$.

We define

\begin{equation}\label{eq:Z_S}
\begin{aligned}
\mathcal{Z}(\omega,\bk_\perp)
&= \ell_\zeta - \bm{\ell}_{\mathrm{d}}^{\top}\mathsf{M}_{\mathrm{dd}}^{-1}\bm{m}_\zeta,\\
\mathcal{S}(\omega,\bk_\perp)
&= f_\zeta - \bm{\ell}_{\mathrm{d}}^{\top}\mathsf{M}_{\mathrm{dd}}^{-1}\mathbf{f}_{\mathrm{d}},
\end{aligned}
\end{equation}

so that the transfer function is

\begin{equation}\label{eq:T}
\mathcal{T}(\omega,\bk_\perp)
=\frac{\zeta}{\mathcal{S}}
=\frac{1}{\mathcal{Z}(\omega,\bk_\perp)}.
\end{equation}

Equivalently, $\mathcal Z$ admits the bordered-determinant (Kreiss–Lopatinskii–Majda) representation \cite{Kreiss70,majda83,metivier2001stability}:

%\footnote{Historically connected to the Lopatinskii–Shapiro condition in boundary-value problems \cite{lopatinskii1953}; we adopt the multi-dimensional shock-stability formulation of \citet{majda83} and \citet{metivier2001stability}.}

\begin{equation}\label{eq:Z_bordered}
\mathcal{Z}(\omega,\bk_\perp)=
\frac{\det\!\begin{bmatrix}
\mathsf{M}_{\mathrm{dd}} & \bm{m}_\zeta\\
\bm{\ell}_{\mathrm{d}}^{\top} & \ell_\zeta
\end{bmatrix}}
{\det \mathsf{M}_{\mathrm{dd}}}\,,
\end{equation}

with the explicit assembly given in Appendix~\ref{app:assembly}.

\begin{figure}
\centering
\includegraphics[width=0.9\columnwidth]{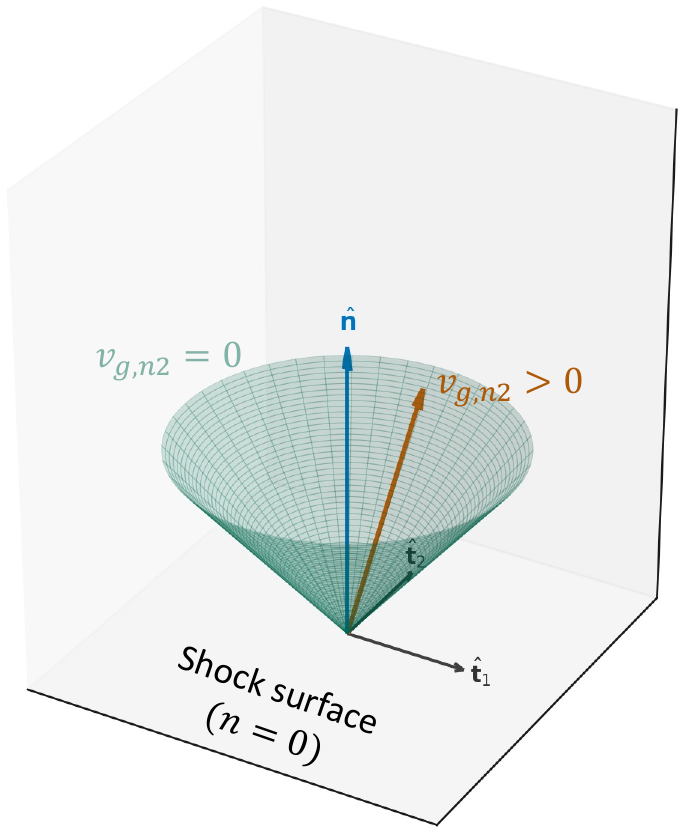}
\caption{Schematic resonance geometry for transmitted downstream fast–magnetosonic modes. For fixed frequency and tangential wavevector $\boldsymbol{k}_\perp$, the fast branch has a locus in $(k_{n2},\boldsymbol{k}_\perp)$ where the normal group speed in the shock frame, $v_{g,n2}$, vanishes. This ``resonance cone'' (green mesh) marks directions where wave packets dwell near the interface and the interfacial impedance $\mathcal{Z}$ is minimized; the closer an orange ray lies to this cone, the stronger the Lorentzian enhancement of the surface response $|\mathcal{T}|^2 = 1/|\mathcal{Z}|^2$.}
\label{fig:rescone2}
\end{figure}

We focus on grazing points of the downstream fast branch where the normal group speed in the shock frame, $v_{g,n2}(\omega,\boldsymbol{k}_\perp)$, vanishes. This condition defines a narrow locus in $(k_{n2},\boldsymbol{k}_\perp)$–space which we refer to as a resonance cone and illustrate schematically in Figure~\ref{fig:rescone2}. At these points the transmitted fast–mode packets propagate nearly parallel to the interface: their residence time near the shock becomes large and the interfacial impedance $\mathcal{Z}$ develops a pronounced minimum. This is an impedance (scattering) resonance of the boundary problem, not a free surface eigenmode.\footnote{$\mathcal Z$ becomes small when the admissible fast branch has $v_{g,n2}\!\to\!0$ in the shock frame; the packet’s residence time at the interface diverges and the interfacial Green’s function peaks. This is a genuine resonance of the transmission problem even though no capillary/Hall–MHD–type bound surface mode is present.}
The elegance of this is due to two straightforward reasons. First, for generic obliquity and parameters the fast mode is the only admissible branch whose normal group speed $v_{g,n2}$ crosses zero at fixed $(\omega,\bk_\perp)$ under the radiation condition. Second, $\zeta$ couples most strongly through compressive (pressure and magnetic-pressure) operators, for which the fast mode dominates the interfacial response. Slow and Alfv\'en branches may contribute, but typically do not produce isolated grazing resonances except under special alignments; their effect is retained in $\mathsf M_{\mathrm{dd}}$ outside the local fast-grazing analysis.

Further details of the bordered-determinant mapping and the linearization near a fast-mode grazing configuration are summarized in Appendix~\ref{app:ZTmapping}. 

The resonance structure of $\mathcal Z$ follows from its dependence on the downstream normal roots through $\mathsf M_{\mathrm{dd}}$. Denote the downstream fast-branch normal group speed by

\begin{equation}\label{eq:vgn_def}
v_{g,n2}(\omega,\bk_\perp)=\bn\!\cdot\!\frac{\partial \omega}{\partial \bk}
\quad\text{subject to}\quad
D_2(\omega-\bk\!\cdot\!\bU_2,\bk)=0.
\end{equation}

A grazing configuration occurs when

\begin{equation}\label{eq:resonance_condition}
v_{g,n2}(\omega,\bk_\perp)=0.
\end{equation}

Differentiation with respect to the fast-mode normal root $k_{n2}$ along the dispersion surface shows that $\mathcal Z$ varies linearly with $v_{g,n2}$ near grazing. Using

\begin{equation}
\label{eq:ratio_D}
\frac{\partial_{k_{n2}} D_2}{\partial_{\omega'} D_2}
= U_{n2}-v_{g,n2},
\qquad
\frac{\partial \omega}{\partial \bk}
= \bU_2-\frac{\partial_{\bk} D_2}{\partial_{\omega'} D_2},
\end{equation}

one finds the coupling coefficient

\begin{equation}\label{eq:C_def}
C(\omega,\bk_\perp)\;\equiv\;
\left.\frac{\partial_{k_{n2}}\mathcal Z(\omega,\bk_\perp)}
{\partial_{k_{n2}}v_{g,n2}(\omega,\bk_\perp)}\right|_{k_{n2}=k_{n2}^{(g)}},
\end{equation}

so that, to first order,

\begin{equation}\label{eq:Z_linear}
\mathcal{Z}(\omega,\bk_\perp)=C(\omega,\bk_\perp)\,v_{g,n2}(\omega,\bk_\perp)+\ii\,\Gamma,
\end{equation}

% where \(\Gamma\) denotes a small effective regularization representing linear sinks (downstream leakage, weak damping, inhomogeneous broadening across \(\bk_\perp\)). 
where $\Gamma$ is a small net regularization (imaginary part) that can represent either damping or growth: $\Gamma>0$ for net damping (dissipation, radiative leakage, effective spectral broadening) and $\Gamma<0$ for net amplification. In particular, $\Gamma$ may include weak shock–front instabilities (e.g., D’Iakov–Kontorovich corrugation \cite{Diakov58a,Diakov58b,kontorovich1958}, and kinetic rippling effects \cite{winske1988}) provided their linear rates are small over the grazing bandwidth and do not alter the admissible-mode set. When such mechanisms are strong or generate additional surface modes, they modify $\mathcal Z$ itself (through $D_2$ and the boundary operators) rather than entering as a simple $i\Gamma$. Practical, parameterized recipes to estimate \(\Gamma\) from along-surface advection, finite root width, and kinetic dissipation are collected in Appendix~\ref{app:Gamma} while any assessment of instabilities is deferred. In the resonance cone of Figure~\ref{fig:rescone2}, $v_{g,n2}\simeq 0$ and the small effective regularization $\Gamma$ therefore sets both the width of the peak response and the lifetime of the coherent corrugation associated with the transmitted fast branch.

%========================
% Figure 0b — Resonance cone
%========================

The surface response is therefore Lorentzian in $v_{g,n2}$,
\begin{equation}\label{eq:T_Lorentz}
|\mathcal{T}(\omega,\bk_\perp)|^2=\frac{1}{|C(\omega,\bk_\perp)|^2\,v_{g,n2}^2(\omega,\bk_\perp)+\Gamma^2},
\end{equation}
which is phenomenologically similar to the grazing resonance in the relativistic regime obtained by \cite{lemoine2016corrugation}.

For clarity, the entropy/contact mode does not generate a grazing-resonance cone in this sense. For the entropy branch $\omega'_{\rm e}=0$ one has $\omega=\bk\!\cdot\!\bU_2$ and therefore $\partial\omega/\partial\bk=\bU_2$, so its normal group speed in the shock frame is

\[
v_{g,n2}^{(\rm e)}=\bn\!\cdot\!\frac{\partial\omega}{\partial\bk}=U_{n2},
\]

which is fixed by the base state and does not pass through $0$ under variation of the downstream normal root at fixed $(\omega,\bk_\perp)$. Entropy fluctuations can still enter the forcing through the boundary operators \cite{lemoine2016corrugation}, but the Lorentzian enhancement discussed here is tied to the transmitted fast branch approaching $v_{g,n2}\simeq 0$.

Because \(\mathcal S\) and \(\mathcal Z\) are separated, the mean-square drive decomposes naturally into Alfv\'enic and compressive contributions,

\begin{equation}\label{eq:S_split}
\begin{aligned}
\avg{|\mathcal{S}(\omega,\bk_\perp)|^2}
&= \mathcal{G}_A(\omega,\bk_\perp)\,E_A(\omega,\bk_\perp) \\
&\quad + \mathcal{G}_C(\omega,\bk_\perp)\,S_C(\omega,\bk_\perp) \\
&\quad + 2\,\Re\!\Big\{\mathcal{G}_{\!AC}(\omega,\bk_\perp)\,
\Phi_{AC}(\omega,\bk_\perp)\Big\},
\end{aligned}
\end{equation}

where \(E_A\) is the upstream Alfv\'enic spectral density projected onto the interface polarization, \(S_C\) is a compressive scalar spectrum associated with pressure and magnetic-pressure couplings, and \(\mathcal G_{A,C}\) are algebraic kernels (explicit quadratic forms of \(\mathbf f_{\mathrm d}\) acted on by \(\bm{\ell}_{\mathrm d}^{\top}\mathsf M_{\mathrm{dd}}^{-1}\)); their reduction can be found in Appendix~\ref{app:weights}. $\Phi_{AC}$ is the Alfv\'enic--compressive cross-spectrum which we set to $0$ (no cross term) when these components are uncorrelated after azimuthal averaging, which is the assumption used in the qualitative scalings obtained here. The general spectrum of corrugations at fixed $(\omega,\bk_\perp)$ is

\begin{equation}\label{eq:zeta_general}
\avg{|\zeta(\omega,\bk_\perp)|^2}
= \frac{\avg{|\mathcal{S}(\omega,\bk_\perp)|^2}}
{|\mathcal{Z}(\omega,\bk_\perp)|^2}.
\end{equation}

Near a fast-mode grazing point, Eq.~\eqref{eq:zeta_general} is therefore a Lorentzian in the grazing variable $v_{g,n2}$. Assuming the drive varies slowly across the resonance width, $\avg{|\mathcal S(\omega,\bk_\perp)|^2}\approx \avg{|\mathcal S(\omega_g,\bk_\perp)|^2}$, we can pull it outside the narrow resonance integral. Evaluating the resonance contribution by integrating across the grazing set using the downstream fast normal root $k_{n2}$ (at fixed $\bk_\perp$) yields the standard Lorentzian integral

\begin{equation*}
\begin{aligned}
&\ \int dk_{n2}\,
\frac{1}{\big|C(\omega_g,\bk_\perp)\big|^2\, v_{g,n2}^2+\Gamma^2} \\
&\qquad =
\frac{\pi}{\big|C(\omega_g,\bk_\perp)\big|\,\Gamma\,\big|\partial_{k_{n2}} v_{g,n2}\big|_{\omega_g}}.
\end{aligned}
\end{equation*}

Thus

\begin{equation}\label{eq:Zk_scaling}
\begin{aligned}
&\ \int dk_{n2}\,\avg{|\zeta(\omega_g,\bk_\perp)|^2} \\
&\ \simeq \frac{\pi}{|C(\bk_\perp)|\,\Gamma}\,
\frac{1}{\bigl|\partial_{k_{n2}} v_{g,n2}\bigr|_{\omega_g}} \\
&\quad \times \Bigl[
\mathcal{G}_A(\bk_\perp)\,E_A(\bk_\perp)
+\mathcal{G}_C(\bk_\perp)\,S_C(\bk_\perp)
\Bigr],
\end{aligned}
\end{equation}

where all quantities are evaluated at the local grazing point $\omega_g(\bk_\perp)$.

In the limit $|U_{n2}|\ll c_{f2}$, the combination of the directional measure of the grazing set and the dwell-time weighting of near-grazing fast packets yields the residence scaling $\propto c_{f2}/|U_{n2}|$ (Appendix~\ref{app:geom_residence}). Using the RH scalings $\Delta\rho\sim\rho_1(r-1)$, $\Delta\bB_\perp\sim(r-1)\bB_{1\perp}$, and $|C|\sim\rho_1 U_{n1}$ (up to geometry), one obtains

\begin{equation}\label{eq:obliquity_scaling}
\begin{aligned}
&\int dk_{n2}\,\avg{|\zeta|^2}(\bk_\perp) \\
&\ \propto\ \frac{(r-1)^2}{\rho_1 U_{n1}\,\Gamma}\,
\frac{c_{f2}}{|U_{n2}|}\,
\Big[\cos^2\theta_{Bn}\,E_A(\bk_\perp)+\sin^2\theta_{Bn}\,S_C(\bk_\perp)\Big],
\end{aligned}
\end{equation}

and with $U_{n2}=U_{n1}/r$ the prefactor reduces to $r\,c_{f2}/(\rho_1 U_{n1}^2\,\Gamma)$, consistent with Eq.~\eqref{eq:kperp_int}. The $\cos^2\theta_{Bn}$ (Alfvénic) and $\sin^2\theta_{Bn}$ (compressive) weights reflect that Alfvénic drive couples through $B_{n1}$ while compressive drive couples through $B_{1\perp}$.\footnote{The net Alfvénic kernel’s $\cos^2\theta_{Bn}$ dependence arises from the full boundary operators (e.g., terms $\propto B_n$ in $\jump{\rho U_n\bU_\perp - B_n\bB_\perp}$ and $\jump{U_n\bB_\perp-\bU_\perp B_n}$) after azimuthal averaging; it does not follow from $\delta U_{n1}^{(A)}$ alone, whose pure projection scales with $|\bB_{1\perp}|$.}

Integrating further over \(\bk_\perp\), the near-grazing directional measure and residence-time effects can be combined into a transparent factorization discussed in Appendix~\ref{app:geom_residence}; in the limit \(|U_{n2}|\ll c_{f2}\) this produces the scaling with \(c_{f2}/|U_{n2}|\) below,

\begin{equation}\label{eq:kperp_int}
\begin{aligned}
&\iint d\omega\,d^2 k_\perp\,\avg{|\zeta|^2} \\
&\ \propto\ \frac{r\,(r-1)^2}{\rho_1 U_{n1}^2\,\Gamma}\,c_{f2}\,
\Big[\cos^2\theta_{Bn}\,E_A^{\mathrm{tot}}+\sin^2\theta_{Bn}\,S_C^{\mathrm{tot}}\Big].
\end{aligned}
\end{equation}

%========================
% Figure 1 — Driver spectrum
%========================
\begin{figure}
\centering
\includegraphics[width=0.9\columnwidth]{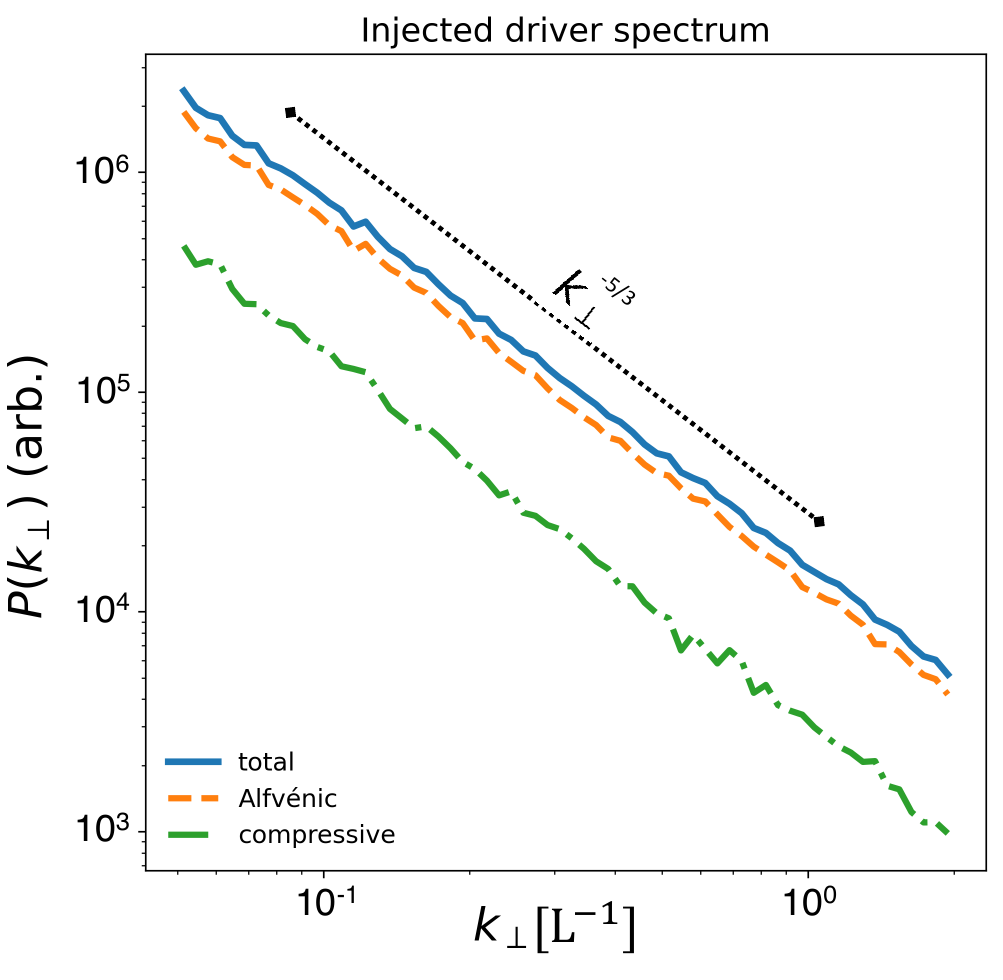}
\caption{Injected broadband driver used in all runs. One–dimensional perpendicular spectrum $P(k_\perp)$ showing the total (solid), Alfv\'enic (dashed), and compressive (dash–dotted) components. A dotted line indicates a $k_\perp^{-5/3}$ guide. Spectra are accumulated from $N=6\times10^4$ upstream samples with a fixed compressive power fraction $\chi_C$. $k_\perp$ in $L^{-1}$ and power in arbitrary units.}
\label{fig:driver}
\end{figure}

while Appendix~\ref{app:geom_residence} documents that the purely geometric directional measure contributes a complementary factor \(\propto c_{f2}^{-1}\) at fixed magnitude \(k\). The upstream spectral choices and polarization content used only to build \(\mathcal S\) are summarized in Appendix~\ref{app:driver}.

\section{Resonant Filtering of Upstream Turbulence and Corrugation Statistics}\label{sec:discussion}

\subsection{Evaluation protocol: base state, injected spectra, and effective regularization}

%========================
% Figure 2 — k–space diagnostics (|k| map, normal-k map, susceptibility)
%========================
\begin{figure*}
\centering
\includegraphics[width=\textwidth]{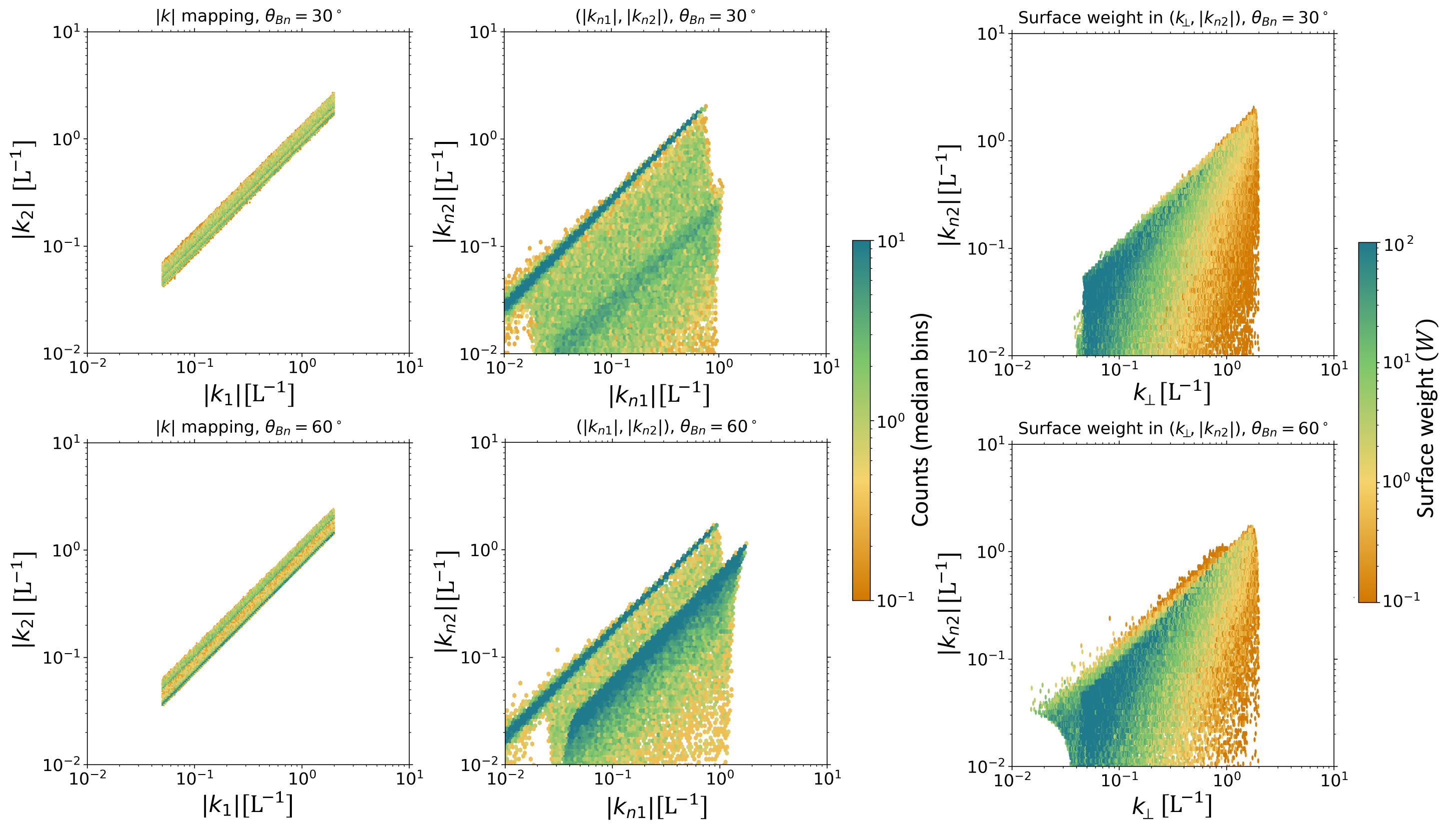}
\caption{Wave–vector diagnostics that relate the upstream driver to the transmitted fast branch and to the surface response. Left column mapping of total wavenumber magnitude $(|k_1|\!\rightarrow\!|k_2|)$ for $\theta_{Bn}=30^\circ$ (top row) and $60^\circ$ (bottom row). Middle column mapping of the normal components $(|k_{n1}|,|k_{n2}|)$ for the same angles and rows. Right column surface weight $W$ in the $(k_{\perp},|k_{n2}|)$ plane (susceptibility map). Color in the left and middle columns shows counts per hexbin, normalized by the median count in each panel. Color in the right column shows $\sum W$ per hexbin, normalized by the panel median. Results use the same $N=6\times10^4$ upstream samples as Figure~\ref{fig:driver} and the base state $\beta=0.1$, $M_f\simeq1.5$.}
\label{fig:kdiagnostics}
\end{figure*}

We evaluate the linear interfacial response for a weak fast shock using the RH base state with $\beta=0.1$, \(\gamma = 5/3\), and $M_f\simeq1.5$. The upstream driver is broadband with a perpendicular spectrum $E(k_\perp)\propto k_\perp^{-5/3}$ and a fixed compressive fraction $\chi_C$ (Fig.~\ref{fig:driver}). In all examples shown here we set $\chi_C=0.2$, so the effective drive at the interface is $(1-\chi_C)\cos^2\theta_{Bn}+\chi_C\sin^2\theta_{Bn}$; the balance of Alfv\'enic and compressive forcing therefore varies strongly with obliquity. For each realization the incident frequency in the shock frame is $\omega=\bk\!\cdot\!\bU_1$. 

The transmitted fast–mode normal wavenumber $k_{n2}$ is obtained from the exact quartic $D(\omega',\bk)=0$ together with the group velocity from the same dispersion. The surface weight used in all diagnostics is

\begin{equation}\label{eq:surf_weight}
\begin{aligned}
&W(\omega,\bk_\perp)\ \\
&\ \propto\ \frac{
(1-\chi_C)\cos^2\theta_{Bn}\,E_A(k_\perp)
+\chi_C\sin^2\theta_{Bn}\,S_C(k_\perp)
}{
\left|C(\omega,\bk_\perp)\right|^2\,v_{g,n2}^2(\omega,\bk_\perp)+\Gamma^2
}.
\end{aligned}
\end{equation}

\(W\) represents the spectral weight with which an upstream fluctuation at \((\omega,\bk_\perp)\) contributes to the surface corrugation. Here $\Gamma$ collects along–surface advection, a finite normal–root width $\Delta k_n$, and a small kinetic term (Appendix~\ref{app:Gamma}). For the qualitative assessment of the model here, we have used \(\Gamma = 5\times10^{-3}\) code units; $\Gamma$ has the same units as $|C|\,v_{g,n2}$; equivalently, $\Gamma/|C|$ is a rate. The surface corrugation $\zeta(t_1,t_2)$ is then synthesized by superposing the selected along–surface modes with random phases.

%========================
% Figure 3 — Real–space slices in (n,t) planes
%========================
\begin{figure*}
\centering
\includegraphics[width=\textwidth]{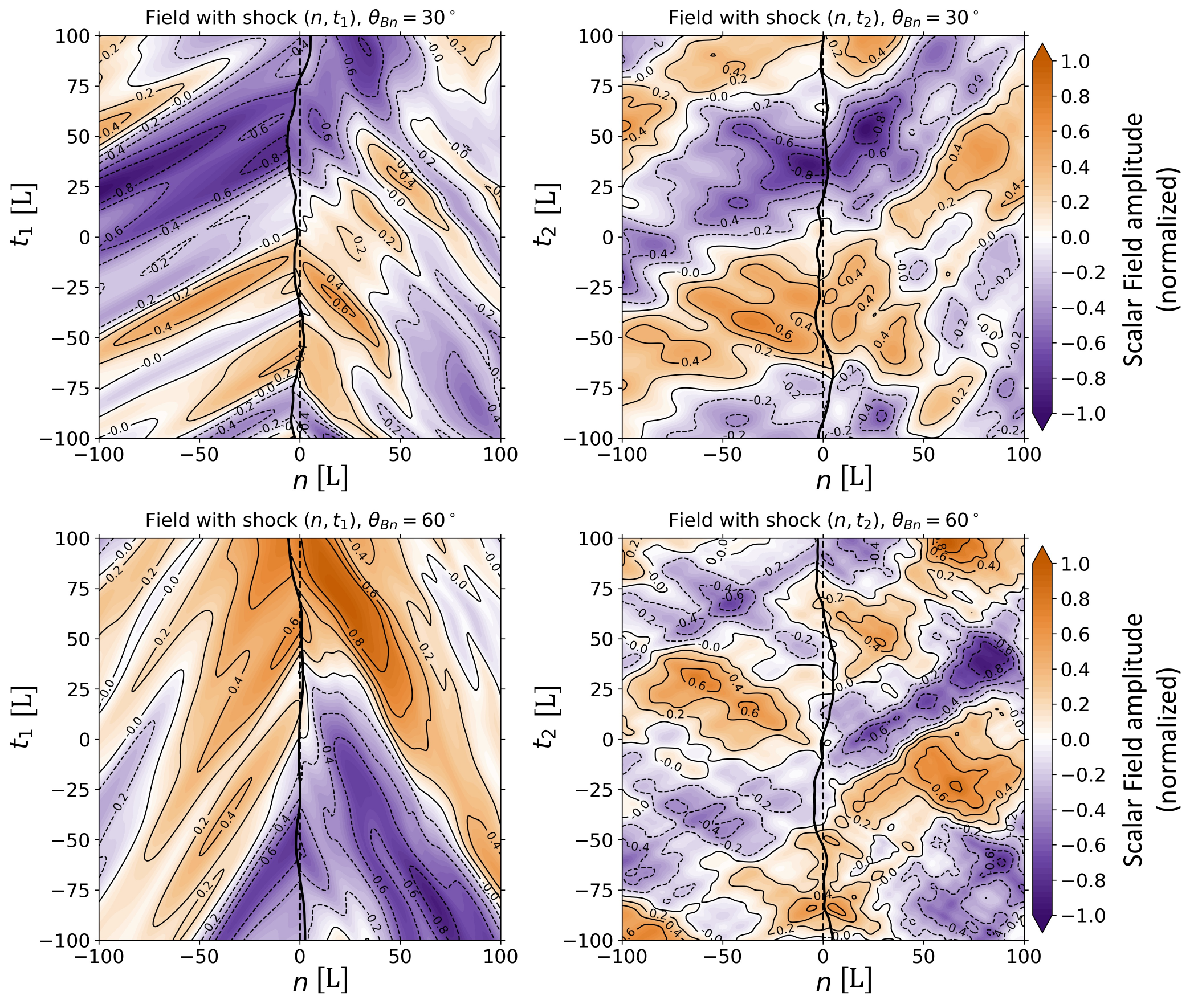}
\caption{Real–space slices of the scalar field in planes that include the shock normal. Panels show normalized scalar–field amplitude in the $(n,t_1)$ left column and $(n,t_2)$ right column for $\theta_{Bn}=30^\circ$ top row and $60^\circ$ bottom row. The solid black curve is the corresponding slice of the synthesized surface $\zeta$ and the vertical dashed line marks $n=0$. Upstream lies to the left of the curve and downstream to the right. Colors are normalized by the absolute maximum within each panel; contour labels indicate the plotted levels.}
\label{fig:realspace}
\end{figure*}

% In the following qualitative assessment, we resolve the interfacial response and the kinematics of the transmitted fast branch; we do not evolve a downstream turbulent cascade.

\subsection{Results: resonance-cone mapping and scaling of surface response}

Figure~\ref{fig:kdiagnostics} shows how the broadband upstream driver is mapped into transmitted downstream fast modes and how the resonance cone of Figure~\ref{fig:rescone2} appears in practice as a weighted ridge in $(k_{\perp},|k_{n2}|)$. In the left column of Figure~\ref{fig:kdiagnostics} the $\left(|k_1|,|k_2|\right)$ cloud lies essentially on the identity, reflecting frequency continuity across the interface together with the weak dispersion of the fast branch at low $\beta$. The faint, nearly parallel rails along this line arise from how the driver populates $(k_\perp,k_\parallel)$ together with logarithmic binning; they are sampling rails, not distinct mode families.

In the middle column of Figure~\ref{fig:kdiagnostics} the normal–component map $\big(|k_{n1}|,|k_{n2}|\big)$ is skewed because phase matching fixes $k_\perp$ while the admissible downstream fast root has a normal component comparable to or slightly larger than the upstream one for most rays. This produces the thin high–intensity rails that lie just above the identity line $|k_{n2}|=|k_{n1}|$. At the same time, near–grazing configurations of the fast branch generate solutions with much smaller $|k_{n2}|$ at fixed $|k_{n1}|$, filling out the broad wedge below the identity line where the downstream normal group speed in the shock frame becomes small. For the more quasi–parallel case ($\theta_{Bn}=30^\circ$) this wedge is relatively diffuse; as the geometry moves toward quasi–perpendicular ($\theta_{Bn}=60^\circ$) it becomes sharper and extends to smaller $|k_{n2}|$. Writing the along–surface wavevector as

\[
\bk_\perp = k_\perp\big(\cos\phi\,\bt_1+\sin\phi\,\bt_2\big),
\]

with $\phi$ the azimuthal angle in the $(t_1,t_2)$ plane and taking $\bB$ in the $(n,t_1)$ plane, the near–grazing condition selects two nearby bands of $\phi$; these two azimuthal families map into the two closely spaced high–intensity rails visible in the $\big(|k_{n1}|,|k_{n2}|\big)$ panels.

The same closely spaced structure is visible as neighboring ridges in the right column of Figure~\ref{fig:kdiagnostics} where $W(k_{\perp},|k_{n2}|)$ is plotted. The susceptibility concentrates on a slanted ridge that is the image of $v_{g,n2}\approx0$ for the normal component while the tangential group speed remains finite. A low–$k_{\perp}$ foot reflects longer along–surface residence time $\tau_{\rm res}\sim L_\parallel/v_{g,\perp 2}$ with $L_\parallel\propto 1/k_{\perp}$, and the high–$k_{\perp}$ taper follows from both the injector cutoff and faster along–surface dephasing. Ridge thickness and contrast track $\Gamma$. Larger $\Delta k_n$ or stronger along–surface dephasing broaden and dim the ridge. Smaller values sharpen it. At very high obliquity the ridge sits at low $k_{\perp}$ and coherence grows even if the ridge broadens at larger $M_f$. At small obliquity the weight spreads to higher $k_{\perp}$ at moderate $|k_{n2}|$ and the low $k_{\perp}$ foot is short, consistent with an Alfv\'enic–leaning driver that does not favor the near–grazing set.

\begin{figure*}
\centering
\includegraphics[width=\textwidth]{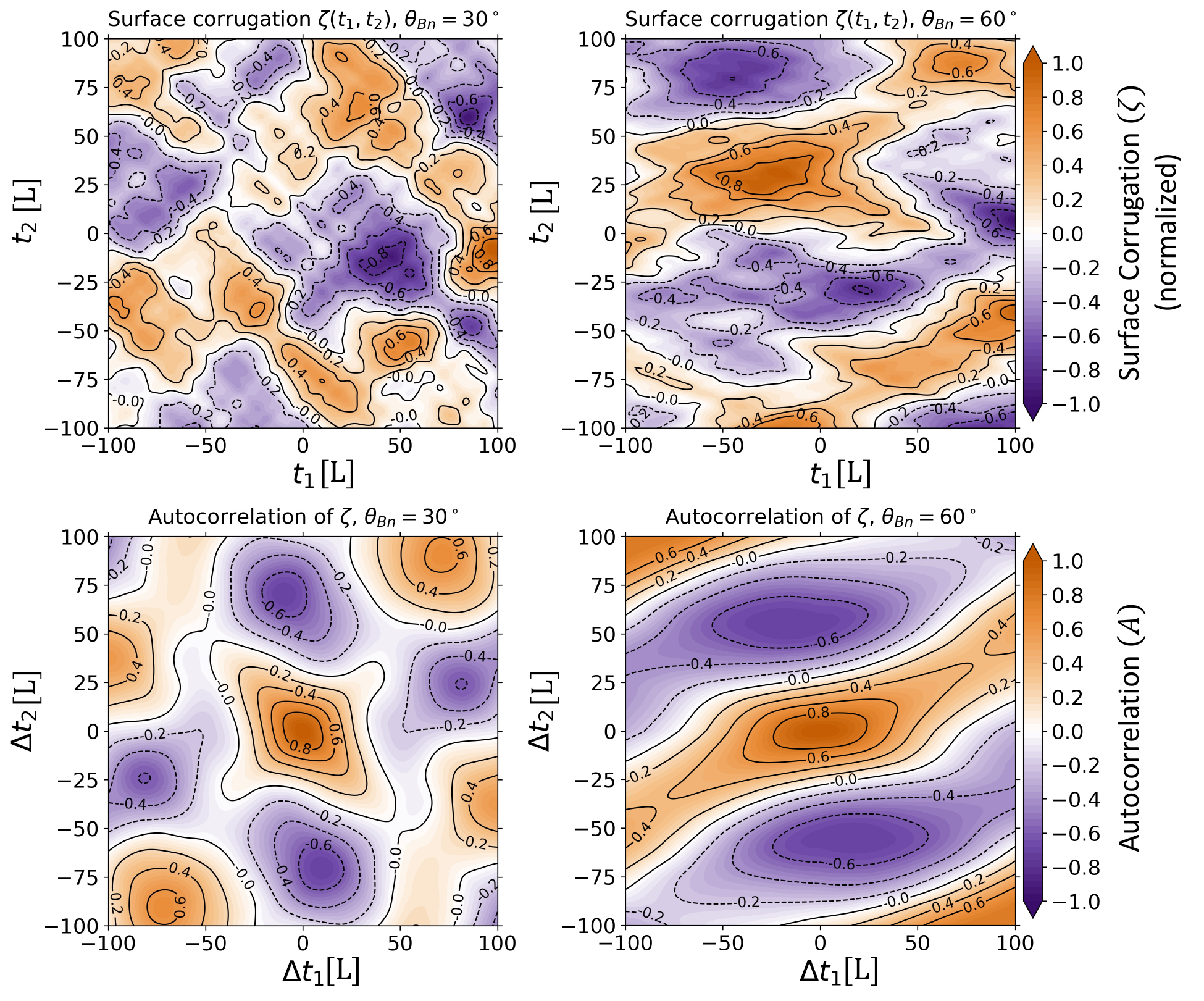}
\caption{Surface corrugation and its autocorrelation. Top panels synthesized surface corrugation $\zeta(t_1,t_2)$ for $\theta_{Bn}=30^\circ$ and $60^\circ$ (normalized to unit peak). Bottom panels corresponding normalized two–dimensional autocorrelation $A(\Delta t_1,\Delta t_2)$ with contours spaced by 0.2. Lags $\Delta t_{1,2}$ are in the same code units as $t_{1,2}$.}
\label{fig:autocorr}
\end{figure*}

These selections propagate cleanly into real space (Figs.~\ref{fig:realspace} and \ref{fig:autocorr}). At $\theta_{Bn}=30^\circ$ the response favors several nearby near–grazing directions. The $(n,t)$ slices display diagonal upstream fronts, a modulated crossing at $n=\zeta$, and downstream bands whose normals tilt toward the surface. At $\theta_{Bn}=60^\circ$ the geometry weights shift power from the Alfv\'enic channel $\propto\cos^2\theta_{Bn}$ toward the compressive channel $\propto\sin^2\theta_{Bn}$. With $\chi_C<1/2$ this reduces the absolute drive but makes it more purely compressive. The near–grazing locus moves, the $\big(|k_{n1}|,|k_{n2}|\big)$ wedge extends further toward small $|k_{n2}|$, the $\zeta$ pattern coarsens into fewer and larger patches, and the along–surface coherence length grows. The total wavenumber mapping remains anchored near the identity in both cases. In Figs.~\ref{fig:realspace} and \ref{fig:autocorr}, the upstream and downstream fields shown together with $\zeta$ visualize the convected approach to the surface, the modulated crossing at $n=\zeta$, and the refraction of transmitted fast waves. They are kinematic reconstructions driven by the interface, not a self–consistent solution for the downstream spectrum.

Figure~\ref{fig:realspace} is the physical check on the $k$–space selection. The interface slice $n=\zeta$ marks the crossing. Upstream diagonal bands approach the surface with angles set by the incident wave vector. Downstream bands bend toward the surface because the transmitted fast branch reduces the normal component of $k$. The spacing of extrema along $n$ gives the local normal wavelength $2\pi/|k_{n2}|$. This spacing grows as rays move toward the near–grazing set, which explains the coarser pattern at $\theta_{Bn}=60^\circ$. The two planes sample directions $t_1$ and $t_2$. With $\bB$ in the $(n,t_1)$ plane the $(n,t_1)$ slice in the left column of Figure~\ref{fig:realspace} carries stronger tilt and longer spacing along the front when the response leans compressive. The $(n,t_2)$ slice in the right column of Figure~\ref{fig:realspace} is less affected. Reading both planes together shows refraction toward the surface and the anisotropy of the selected along–front content. The slices also make clear that the surface controls the downstream phase. The phase is continuous at $n=\zeta$ and the change in $k_{n}$ across the interface sets the visible rotation of the wavefronts. In the quasi–perpendicular case the downstream normal spacing increases as rays approach the near–grazing set, which reads as coarser bands. In the small–obliquity case the spacing is finer because the selected $|k_{n2}|$ is larger on average. All fields and $\zeta$ maps are normalized for display, so morphology and coherence can be read directly while absolute amplitudes should be taken from $\langle|\zeta|^2\rangle$ before normalization.

%========================
% Figure 4 — Surface corrugation and autocorrelation
%========================

We summarize the real–space morphology with the two–dimensional autocorrelation \cite{dudokdewit2003,Bendat11}
\begin{equation} \label{eq:autocorr}
    A(\Delta t_1,\Delta t_2)\ \equiv\ \frac{\big\langle \zeta(t_1,t_2)\,\zeta(t_1+\Delta t_1,\,t_2+\Delta t_2)\big\rangle}{\big\langle \zeta^2 \big\rangle},
\end{equation}
computed on the same grid as \(\zeta\). Here \(A\) is dimensionless and the lags \(\Delta t_{1,2}\) carry the same code length units as \(t_{1,2}\). In the bottom panels of Figure~\ref{fig:autocorr} the major axis of the central lobe is parallel to the dominant direction of corrugation in \((t_1,t_2)\). Its width along that axis is the practical measure of the along–front coherence length \(L_\parallel\). A narrower lobe implies larger selected \(k_{\perp}\). A broader lobe implies stronger low \(k_{\perp}\) weight and longer residence. Fine shoulders or weak secondary ridges at nearly fixed \(|\Delta t|\) arise when several nearby \(k_{\perp}\) bands are selected by the near–grazing ridge in $W(k_{\perp},|k_{n2}|)$. In our calculations, the \(30^\circ\) map has a compact core with weak shoulders and only modest anisotropy. The \(60^\circ\) map has a broader core with diminished shoulders, consistent with a shift of weight toward lower $k_{\perp}$ and a longer residence factor $c_{f2}/|U_{n2}|$.

In the bottom panels of Figure~\ref{fig:autocorr} the orientation of the major axis gives the direction of corrugation in $(t_1,t_2)$ and matches the near–grazing families selected in the right column of Figure~\ref{fig:kdiagnostics}. The half–maximum width of the central lobe along that axis is our practical $L_\parallel$ and tracks the inverse of the selected $k_{\perp}$ bandwidth in $W(k_{\perp},|k_{n2}|)$. The minor–axis width gives $L_\perp$ and the ratio $L_\parallel/L_\perp$ measures anisotropy of the surface pattern. Shoulders at nearly fixed $|\Delta t|$ appear when the ridge selects more than one nearby $k_{\perp}$ band. In our runs the case with $\theta_{Bn}=60^\circ$ has larger $L_\parallel$ and larger anisotropy than the case with $\theta_{Bn}=30^\circ$. This is consistent with the shift of weight toward lower $k_{\perp}$ and the longer residence that we read in the right column of Figure~\ref{fig:kdiagnostics} and that we see as coarser patches in the bottom row of Figure~\ref{fig:realspace}.

For quasi–perpendicular geometry the selected bands shift to lower $k_{\perp}$ and the coherence grows. These patterns are consistent with nonplanar fronts in quasi–perpendicular shocks at 1~AU, where multispacecraft measurements report local radii of curvature from $3\times10^{5}$ to $10^{7}$~km with an average near $3.5\times10^{6}$~km, and with bay–like source regions that generate multiple drifting lanes in type~II emission \cite{Neugebauer05,Bale99}. If scaled to the size of a SNR, the \(L_\parallel/L_\perp\) anisotropy in Fig.~\ref{fig:autocorr} (bottom right) would appear as a train of coherent, elongated X-ray stripes whose long axis follows the dominant corrugation direction and whose spacing reflects the along-front coherence \(L_\parallel\). Tycho shows such stripes with characteristic separations \(\sim10^{10}\,\mathrm{km}\), consistent with this mapping \cite{eriksen2011evidence}.

Changes in $M_f$ and changes in obliquity act in different ways. Larger $M_f$ increases $r$ but also increases the effective $\Gamma$, while a move toward perpendicular geometry shifts selection to lower $k_{\perp}$ and lengthens residence. In the quasi–perpendicular case the geometric shift can dominate and coherence can grow even if the ridge broadens.

While we do not construct the full boundary system or the Schur determinants here, the amplitude and coherence trends follow directly from the impedance. At fixed $\bk_\perp$,

\begin{equation}
  \begin{aligned}
    &\big\langle|\zeta|^2\big\rangle(\bk_\perp)\ \\
    &\propto\ 
    \frac{(r-1)^2}{\rho_1U_{n1}\,\Gamma}\,
    \frac{c_{f2}}{|U_{n2}|}\,
    \Bigg[\begin{gathered}
      (1-\chi_C)\cos^2\theta_{Bn}\,E_A(\bk_\perp) \\
      + \chi_C\sin^2\theta_{Bn}\,S_C(\bk_\perp)
    \end{gathered}\Bigg],
  \end{aligned}
\end{equation}

which shows that, at the level of this analytic scaling, the characteristic amplitude at fixed $\bk_\perp$ is boosted by the explicit compression factor $(r-1)^2$ and by the residence factor $c_{f2}/|U_{n2}|$, and is reduced by the inflow $\rho_1 U_{n1}$ and by the effective damping $\Gamma$, while the relative Alfv\'enic and compressive weights follow the $\cos^2\theta_{Bn}$ and $\sin^2\theta_{Bn}$ factors in Eq.~\eqref{eq:obliquity_scaling}. 

In practice, $r$ also enters indirectly through $U_{n2}=U_{n1}/r$, through $c_{f2}$, and through $\Gamma$, so the net dependence on compression reflects a balance of these contributions rather than $(r-1)^2$ alone. After integrating over $\bk_\perp$ the purely geometric directional measure contributes an additional factor, yielding the scaling in Eq.~\eqref{eq:kperp_int}. Changing the fast Mach number $M_f$ at fixed $\beta$ modifies both $r$ and $U_{n1}$, and typically also $\Gamma$ through changes in residence and radiative leakage, so this scaling does not imply a universal monotonic dependence of corrugation power on $M_f$; for weak shocks with $\beta\simeq0.1$ and $M_f\gtrsim1$ the analytic balance suggests that modest increases in $M_f$ can lead to a net growth of corrugation power when the increase in $r$ dominates the additional loading of the denominator by $|U_{n1}|$, but this behaviour is parameter dependent rather than generic. Changing $\beta$ modifies both $c_{f2}$ and the jump in the normal flow, so the residence factor $c_{f2}/|U_{n2}|$ is not monotone in $\beta$; within the narrow range of $\beta$ we explore its effect is secondary to those of obliquity and compression. Varying $\chi_C$ shifts the preferred obliquity through the $\cos^2\theta_{Bn}$ and $\sin^2\theta_{Bn}$ weights, moving the response between more oblique Alfv\'enic--leaning sectors and more quasi--perpendicular compressive--leaning sectors. The along--surface speed of corrugations in the shock frame is $v_{\rm corug}\simeq U_{2\perp}+v'_{g,\perp 2}$; in the downstream fluid frame $v'_{g,\perp 2}\approx c_{f2}\sin\theta_g$ with $\theta_g$ defined by $v_{g,n2}=0$.

The spectrum of corrugations at fixed $\bk_\perp$ is the upstream spectrum filtered by the shock impedance. If the upstream driving is Kolmogorov as in Fig.~\ref{fig:driver} and if the effective damping $\Gamma$ is roughly constant across the band the corrugations inherit the $-5/3$ slope. If damping is dominated by along front advection so that $\Gamma\propto k_\perp$ small scales are preferentially drained and the spectrum steepens toward $-8/3$. The advective contribution to the regularization can be written as a rate $\nu_{\rm adv}\propto v_{g,\perp 2}/L_\parallel\simeq c_{f2}\sin\theta_g/L_\parallel$, and we absorb $|C|$ into $\Gamma$ so that $\Gamma_{\rm adv}\sim |C|\,\nu_{\rm adv}$. So, larger lateral group speed increases \(\Gamma\) and preferentially drains small scales, lowering the amplitude of corrugations at high \(k_\perp\).

As a caveat, our evaluation resolves the surface displacement and enforces the continuity conditions at \(n=\zeta\) while keeping the mean jump fixed at the planar RH state. The broadband driver is treated in linear superposition and wave averaged second order stresses are not used to update the jump parameters, so the formulation applies for small slopes \(\varepsilon\equiv k_\perp|\zeta|\ll1\) and for fluctuation energy densities that remain small compared with the ram and magnetic pressures. Possible nonlinear evolution of these near grazing packets, for example in a Burgers style reduction along the front \cite{Whitham74}, is an interesting extension but lies outside the scope of the present work. 

If fluctuation pressure or momentum flux becomes appreciable \cite{Golan25}, the jump conditions and the stationarity of the front must be solved self-consistently \cite[e.g.][]{Vainio99,Gedalin23}; in that regime the effective \(\Gamma\) and the location of the near grazing ridge depend on the modified base state and may shift. Furthermore, any change to the equation of state also requires reassessing the stability and admissibility of the new solution in the sense of Lax \cite{blokhin1999stability} \footnote{Although, it has been demonstrated that fast-magnetosonic shocks are mostly stable even when \(\gamma\sim 1\) \cite{sotnikov2020collision}.}. 

%The non-linear evolution of these modes For amplitudes approaching $\mathcal{O}(1)$ the local normal tilts significantly and Lax admissibility should be re-evaluated \cite{Akhiezer75}. 

\section{Particle response and implications for DSA}
\label{sec:accel_implications}

Particle acceleration is a central aspect of shock physics \cite{Malkov01}. Here we restrict attention to how a corrugated surface modulates the local obliquity and, through that, any obliquity–sensitive injection. A full DSA treatment for corrugated shocks is deferred.

With the moving surface at $n=\zeta(\boldsymbol{x}_\perp,t)$ and $|\nabla_{\!\perp}\zeta|\ll1$, the perturbed unit normal is $\hat{\boldsymbol n}\simeq \bn-\nabla_{\!\perp}\zeta$, hence

\begin{equation}
\label{eq:accel_dtheta}
\delta\theta_{Bn}
\simeq
\frac{\bB_{1\perp}\!\cdot\!\nabla_{\!\perp}\zeta}{|\bB_1|\,\sin\theta_{Bn}},
\quad
\delta\theta_{Bn,\bk_\perp}
=
\ii\,\frac{\bB_{1\perp}\!\cdot\!\bk_\perp}{|\bB_1|\,\sin\theta_{Bn}}\,
\zeta_{\bk_\perp}.
\end{equation}

For a single Fourier mode the obliquity modulation is governed by the geometric steepness $\varepsilon_{\bk_\perp}\!=\!k_\perp|\zeta_{\bk_\perp}|\!\ll\!1$:

\begin{equation}
\label{eq:accel_dtheta_eps}
\big|\delta\theta_{Bn,\bk_\perp}\big|
= k_\perp\,|\zeta_{\bk_\perp}|\,|\cos\phi|
\equiv \varepsilon_{\bk_\perp}\,|\cos\phi| ,
\end{equation}
where $\phi$ is the angle between $\bk_\perp$ and $\bB_{1\perp}$. Note that any explicit $\theta_{Bn}$–dependence cancels in \eqref{eq:accel_dtheta_eps} because $|\bB_{1\perp}|=|\bB_1|\sin\theta_{Bn}$ at this order.

Let $\Psi(\theta_{Bn})$ be the injected fraction of a thermal population. Linearizing,
\begin{equation}
\label{eq:accel_deltaeta}
\delta\Psi_{\bk_\perp}
=
\Psi'(\theta_{Bn})\,\delta\theta_{Bn,\bk_\perp}
=
\ii\,\Psi'(\theta_{Bn})\,
\frac{\bB_{1\perp}\!\cdot\!\bk_\perp}{|\bB_1|\,\sin\theta_{Bn}}\,
\zeta_{\bk_\perp}.
\end{equation}
The sign of $\Psi'$ is species/environment dependent, but the geometry driving $\delta\theta_{Bn}$ is common. Hence the fractional modulation scales as
\begin{equation}
\label{eq:accel_frac_mod}
\frac{|\delta\Psi_{\bk_\perp}|}{\Psi(\theta_{Bn})}
\simeq
\left|\frac{\Psi'(\theta_{Bn})}{\Psi(\theta_{Bn})}\right|
\varepsilon_{\bk_\perp}\times\mathcal{O}(1).
\end{equation}
The steepness $\varepsilon_{\bk_\perp}$ is set by the interfacial impedance scalings in Eqs.~\eqref{eq:obliquity_scaling}–\eqref{eq:kperp_int}.

If an energetic population contributes to the normal stress, a minimal single–pole closure maps an injection perturbation into an extra compressive drive with transfer

\begin{equation}\label{eq:accel_Tcr}
\begin{aligned}
\mathcal{T}_{\rm cr}(\omega,\bk_\perp)
&= \frac{t_{\rm acc}}{\,1-\ii\omega t_{\rm acc}+\mathcal{K}\, t_{\rm acc}\, k_\perp^2\,},\\
t_{\rm acc}
&\approx \chi_{\mathrm{acc}}\,\frac{\kappa}{U_{n1}^2},\\
\chi_{\mathrm{acc}}
&= \frac{3\,r(r+1)}{r-1},\qquad \kappa_1=\kappa_2=\kappa .
\end{aligned}
\end{equation}

Here \(\kappa\) is the (scalar) particle spatial diffusion coefficient that sets the planar DSA time, while \(\mathcal{K}\) is an effective tangential diffusivity in the along--surface reaction--diffusion closure (Appendix~\ref{app:Tcr_derivation}). Anisotropic transport may be incorporated by replacing \(\kappa\) with the appropriate normal/upstream--downstream projections in \(t_{\rm acc}\), and \(\mathcal{K}\) with the appropriate along--front projection in the \(k_\perp^2\) term.

The coupling is therefore given by

\begin{equation}
\label{eq:accel_Scr}
\begin{aligned}
& S_{\rm cr}=\Lambda_{\rm cr}(\bk_\perp)\,\mathcal{T}_{\rm cr}\,\zeta,\\
&\Lambda_{\rm cr}
=
\ii\,\mathcal A_{\rm cr}\,\Psi'(\theta_{Bn})\,
\frac{\bB_{1\perp}\!\cdot\!\bk_\perp}{|\bB_1|\,\sin\theta_{Bn}},  
\end{aligned}
\end{equation}

where $\mathcal A_{\rm cr}$ collects the conversion from injected–fraction perturbation to a normal–stress perturbation (chosen so that $\Lambda_{\rm cr}\mathcal T_{\rm cr}$ carries the units of $\mathcal Z$; the value depends on the normalization in Appendix~\ref{app:Tcr_derivation}). Because the interface equation is scalar, this feedback renormalizes the impedance,

\begin{equation}
\label{eq:accel_Zeff}
\big(C\,v_{g,n2}+\ii\Gamma-\Lambda_{\rm cr}\mathcal{T}_{\rm cr}\big)\,\zeta
= \mathcal S,
\end{equation}

with $\mathcal S$ the upstream drive of Eq.~\eqref{eq:Z_S} evaluated without the feedback term. To leading order the damping is shifted as

\[
\Gamma\ \to\ \Gamma_{\rm eff}=\Gamma-\Im\!\big[\Lambda_{\rm cr}\mathcal{T}_{\rm cr}\big],
\]

while $\Re[\Lambda_{\rm cr}\mathcal{T}_{\rm cr}]$ shifts the reactive (stiffness) part $C v_{g,n2}$. A linear surface–feedback instability requires $\Gamma_{\rm eff}<0$, i.e. $\Im[\Lambda_{\rm cr}\mathcal{T}_{\rm cr}]>\Gamma$. The sign of $\Im[\Lambda_{\rm cr}\mathcal T_{\rm cr}]$ depends on $\Psi'(\theta_{Bn})$ and on the azimuthal alignment $\mathrm{sgn}(\bB_{1\perp}\!\cdot\!\bk_\perp)$; growth demands that the azimuthally averaged contribution overcome $\Gamma$.

Evaluating on the drift $\omega\!\simeq\! k_\perp v_{\rm corug}$ of the corrugations and equating a quarter–period (an $\mathcal O(1)$ choice) to the acceleration time defines a characteristic matching scale,
\begin{equation}\label{eq:accel_kstar}
\begin{aligned}
k_\ast &\simeq \frac{\pi}{2\,\chi_{\mathrm{acc}}}\,
\frac{U_{n1}^2}{\kappa\,v_{\rm corug}},\\[4pt]
\lambda_\parallel &\simeq \frac{2\pi}{k_\ast}
= \frac{4\,\chi_{\mathrm{acc}}\,\kappa}{U_{n1}^2}\,v_{\rm corug},\\[4pt]
\Delta t_{\rm patch}
&=\frac{2\pi}{k_\ast v_{\rm corug}}
=\frac{4\,\chi_{\mathrm{acc}}\,\kappa}{U_{n1}^2}.
\end{aligned}
\end{equation}

Diffusion further suppresses the response at large $k_\perp$ through the factor $(1+\mathcal{K}\,t_{\rm acc}k_\perp^2)$ in \eqref{eq:accel_Tcr}; hence the peak response is biased toward the smallest available $k_\perp$ when diffusion dominates.

Equations~\eqref{eq:obliquity_scaling}–\eqref{eq:kperp_int} imply that quasi–perpendicular shocks, which carry larger $\sin^2\theta_{Bn}$ weight and longer residence $c_{f2}/|U_{n2}|$, preferentially select lower $k_\perp$. Consequently $L_\parallel$ increases. Whether the steepness $\varepsilon_{\bk_\perp}=k_\perp|\zeta_{\bk_\perp}|$ increases at those selected scales depends on the $k_\perp$–dependence of $\Gamma$: if advective regularization dominates ($\Gamma_{\rm adv}\propto v_{g,\perp}\,k_\perp$, with $v_{g,\perp}\equiv \bk_\perp\!\cdot\!(\partial\omega/\partial\bk)/k_\perp$ the along–front group speed of the fast branch), then $|\zeta_{\bk_\perp}|\propto 1/k_\perp$ near grazing and $\varepsilon_{\bk_\perp}$ is roughly constant; if leakage/kinetic terms dominate so that $\Gamma$ is weakly $k_\perp$–dependent, $|\zeta_{\bk_\perp}|$ can grow fast enough that $\varepsilon_{\bk_\perp}$ increases as selection shifts to lower $k_\perp$. These trends are independent of the sign of $\Psi'$ and apply to ions and electrons through the common geometric factor $\varepsilon_{\bk_\perp}$.

Within this linear mesoscale framework, corrugation–driven injection primarily rescales the normalization of accelerated populations at fixed $r$; spectral slopes remain unchanged unless energetic feedback is strong enough to alter $\Gamma$ and the RH base state via Eq.~\eqref{eq:accel_Zeff}. Two operational predictions follow from Eqs.~\eqref{eq:accel_Tcr}–\eqref{eq:accel_kstar}: the along–front spacing of hot spots scales as $\lambda_\parallel\propto v_{\rm corug}$, while the local recurrence time is $\Delta t_{\rm patch}\propto \kappa/U_{n1}^2$.

\section{Conclusions}\label{sec:conclusions}

In this study, we present a novel interfacial formulation within linear MHD that offers an observation-ready map from upstream turbulence and geometry to surface corrugation without modifying the underlying conservation laws. Several conclusions can be made as follows:

\begin{enumerate}
\item Fast-magnetosonic shocks act as directional filters. Broadband upstream power focuses into transmitted fast modes that approach near-grazing, producing mesoscale corrugations that drift along the front in the shock frame at approximately \(v_{\rm corug}\simeq U_{2\perp}+c_{f2}\sin\theta_g\), where \(c_{f2}\sin\theta_g\) is the lateral fast-mode group speed in the downstream frame.

\item The spectrum of corrugations follows the upstream spectrum of turbulence. With weak, roughly scale–independent damping, the corrugations inherit the driver dependence. Stronger damping and along–front advection (so that $\Gamma\propto k_\perp$) preferentially drain smaller scales and, over the range where the effective impedance varies slowly with $k_\perp$, steepen the spectrum by roughly one power in $k_\perp$ (from $-5/3$ toward $\simeq -8/3$), consistent with recent turbulence diagnostics in young SNRs \cite{petruk2025properties}.

\item The amplitude and coherence of the corrugations are set by the interface response. At fixed scales, the amplitude grows with compression and is limited by the normal inflow and weak damping. Increasing $M_f$ strengthens the response, while very large $M_f$ can shorten coherence unless the selection shifts to larger scales.

\item Obliquity and the mix of driver modes set the selection. Alfv\'enic fluctuations at small obliquity favor multiple co-moving bands with shorter along-front coherence. Compressive fluctuations toward perpendicular geometry favor fewer directions with coarser, longer-lived patches and larger anisotropy.

\item Coherence grows with $c_{f2}/|U_{n2}|$ and with a narrower selected $k_{\perp}$ band, provided that the effective damping rate $\Gamma$ remains modest and does not vary too strongly across that band. Because $\beta$ changes both $c_{f2}$ and the jump, the trends are not monotonic. Within this weak–damping picture we expect corrugations to be long and coherent in the low–$\beta$ corona and patchier near 1\,AU. Coherence is shortest at high–$M_f$ bow shocks where $\Gamma$ is expected to be stronger and more scale dependent.

\item The slopes of the corrugations also perturb the local obliquity and modulate particle injection on top of whatever baseline efficiency is set by the available seed particles and Mach-number--dependent shock microphysics. In our model, keeping this baseline fixed, the injected fraction increases with geometric steepness $k_\perp|\zeta|$ and predominantly changes the normalization of the accelerated population rather than its spectral slope.
\end{enumerate}

Across heliospheric and astrophysical settings, a shock moving through a turbulent medium organizes broadband fluctuations into coherent corrugations that slide along the surface. This process is fundamentally different from instability-driven corrugations of fast-magnetosonic shocks and is compatible with well-known results showing that collisionless fast-magnetosonic shocks are linearly stable against MHD corrugational instability \cite{Diakov58a,Diakov58b,kontorovich1958} under many realistic conditions \cite[e.g.][]{Zank87,blokhin1999stability}. In this view, turbulence-driven corrugation provides a minimal, self-consistent baseline description, while more specific kinetic or geometric effects can be added as required by a given system. In multi-spacecraft observations at interplanetary shocks and at the Earth bow shock, the outcome appears as changes in the local shock normal on scales larger than ion scales, limited by the correlation length of the upstream forcing and by damping at the shock \cite{Neugebauer05,kajdivc2019first}.

When the upstream forcing is coherent across a sector, the same filtering at bow shocks across the heliosphere can naturally contribute to the slow ``breathing'' with periods of minutes and excursions of a few hundred kilometres \cite[e.g.][]{Huterer97,madanian25,cheng2025bow}. In such cases the global geometry remains nearly fixed while the local normal tilts by only a few degrees \cite{Horbury02,meziane2014shape}. Those tilts are sufficient to change \(M_n = M\cos\theta_{Bn}\) and to move patches between more parallel and more perpendicular character, producing multi-crossings with nearly steady average normals and episodic changes in reflected-ion activity and wave power. In our framework, the amplitude and persistence follow from how long disturbances reside along the surface compared with how quickly they are drained, and from how coherent the driver is. The period and apparent drift then reflect the along-front scale selected by the interface response, so breathing becomes a natural and interpretable outcome of the shock’s interaction with turbulence, even in the absence of additional instabilities.

In the low corona, similar surface corrugation can help explain rapidly evolving type~II radio sources and fine structure in dynamic spectra, including multiple bright bands that move along the front \cite{Morosan25,Magdalenic20}. Recent studies of interplanetary shocks show that synchrotron features can separate into lanes with differing apparent drifts when relativistic electrons are preferentially confined to distinct corrugations \cite{Jebaraj24b,Wilson25}, a behaviour that is naturally accommodated within our framework. In SNR shocks, an analogous organization is consistent with elongated stripes or smaller patches whose orientation follows the direction of the corrugations, while their spacing and persistence reflect the surface coherence and the damping \cite{eriksen2011evidence,Raymond20}. Corrugations that tilt the local obliquity \(\theta_{Bn}\) create pockets of enhanced injection and modulate synchrotron brightness along the front, providing observables that could, in principle, be used together with complementary constraints and detailed forward modelling to infer the jump parameters, obliquity, and the relative Alfv\'enic/compressive content of the upstream turbulence. 

Beyond morphology, the interfacial framework clarifies how energy is redistributed at the boundary. Near-grazing transmission redirects part of the downstream fluctuation energy into along-front propagation, yielding an anisotropic flux tied to the selected directions of the corrugations and their coherence. In practice, this acts as an effective boundary condition for turbulence and particle transport across the shock, with the degree of along-front redirection set by geometry and compression. A full redistribution budget, including any nonlinear cascade downstream, lies beyond the present linear treatment.

At shocks such as young SNRs and strong heliospheric shocks energetic particles and reflected ions can feed energy back into the shock face. In our framework the shock is a resonant filter rather than a self-excited oscillator; amplification appears only if this feedback outweighs the net damping at the surface. When it does, the preference is for longer along-front wavelengths: larger corrugations give particles time to react within a cycle, while diffusion and other smoothing processes drain the smaller corrugations. Stronger compression enhances the coupling, and oblique magnetic geometry sets both the sign and the strength of the effect. If kinetic feedback becomes strong enough to change which waves are supported at the interface, the effect lies beyond the present linear-MHD closure and should be captured by updating the interfacial impedance rather than by a simple effective damping term.

Finally, a corrugated, drifting boundary makes the effective diffusion tensor anisotropic and time-dependent even when microscopic scattering is unchanged, thereby reshaping particle transport at the shock. Along the front, long coherent stretches increase residence, steady the source, and raise escape thresholds; short, broken stretches yield bursty injection and patchy acceleration. Across the front, slow wandering of the shock normal permits occasional cross-field (and non-diffusive) steps that a fixed plane would suppress. As a result, coarse-grained mean free paths, escape probabilities, and residence times depend on the coherence length, drift speed, and damping that set the corrugation pattern—providing concrete inputs for transport models near shocks.

With this linear baseline, one can quantify how rippling modifies particle injection and transport, revisit turbulence transport closures with anisotropic and time-dependent boundary forcing, and connect emissivity maps to parameters that describe both the shock and its environment. The clearest impacts are expected in effective diffusion near shocks, in escape and residence diagnostics used in acceleration models, and in the predicted variability of radio and X-ray emission.

\begin{acknowledgments}
I.C.J, M.M, and V.V.K dedicate this work to the memory of their dear friend and colleague, Michael (Misha) Balikhin.
This research was supported by the International Space Science Institute (ISSI) in Bern through ISSI International Team project No.~575, ``\textit{Collisionless Shock as a Self-Regulatory System}''.
I.C.J. acknowledges support from the Research Council of Finland (X-Scale, grant No.~371569). 
I.C.J and V.V.K acknowledge support from ISSI's ``Visiting Scientist Program''.
M. M.  was supported by NASA ATP 80NSSC24K0774 and Fermi 80NSSC25K7346 grants. 
N.D. is grateful for support by the Research Council of Finland (SHOCKSEE, grant No.\ 346902, and AIPAD, grant No.\ 368509). 
\end{acknowledgments}

% \begin{contribution}
% I.C.J formulated the problem and conceptualized the theoretical framework with assistance from M.M and V.V.K. The mathematical solution was obtained by I.C.J and independently re-derived by N.W. The feasibility of the model was evaluated by M.M, N.W, J.P, and R.V. The practical applications of the model in both astrophysical and heliospheric context was discussed together with M.M, and N.D. 
% \end{contribution}

\newpage

\appendix

\section{Magnetosonic Polarization in the \texorpdfstring{$(\bk,\bB)$}{(k,B)} Plane}\label{app:pol}

We decompose vectors in the \((\bk,\bB)\) plane with \(\hat{\bk}=\bk/k\), \(\hat{\bm\xi}\) the unit vector in that plane orthogonal to \(\hat{\bk}\), and \(\hat{\bm e}_A=\hat{\bk}\times\hat{\bm\xi}\). For a compressive disturbance \cite[][]{Kadomtsev66},

\begin{equation}\label{app:eq:pol_ansatz}
\delta\bU=\alpha\,\hat{\bk}+\eta\,\hat{\bm\xi},\qquad \delta\bU\!\cdot\!\hat{\bm e}_A=0.
\end{equation}

Linearized relations (uniform background) are \cite[][]{Freidberg14}
\begin{equation}\label{app:eq:cont}
-\,\ii\omega'\,\delta\rho+\ii\rho\,\bk\!\cdot\!\delta\bU=0
\ \Rightarrow\ 
\delta\rho=\frac{\rho k}{\omega'}\,\alpha ,
\end{equation}

\begin{equation}\label{app:eq:ind}
\begin{aligned}
-\,\ii\omega'\,\delta\bB=\ii\,\bk\times(\delta\bU\times\bB)
\ \\\Rightarrow\ 
\delta\bB=\frac{1}{\omega'}\Big[\bB\,(\bk\!\cdot\!\delta\bU)
-\delta\bU\,(\bk\!\cdot\!\bB)\Big],    
\end{aligned}
\end{equation}

\begin{equation}\label{app:eq:mom}
\begin{aligned}
& -\,\ii\omega'\rho\,\delta\bU
= -\,\ii\bk\,\delta p
+ \ii\big[(\bk\times\delta\bB)\times\bB\big],\\ 
&\delta p= c_{\mathrm s}^2\delta\rho,\ \\
&\ c_{\mathrm s}^2=\gamma p/\rho .   
\end{aligned}
\end{equation}

Using \(\bk\!\cdot\!\delta\bB=0\), one finds \(\delta\bB\parallel\hat{\bm\xi}\) with

\begin{equation}\label{app:eq:deltaBxi}
\delta\bB=\frac{k}{\omega'}\big(\alpha\,B_\perp-\eta\,B_\parallel\big)\,\hat{\bm\xi},
\qquad
B_\parallel=\bB\!\cdot\!\hat{\bk},\ \ 
B_\perp^2=B^2-B_\parallel^2 .
\end{equation}

Projecting Eq.~\eqref{app:eq:mom} onto \(\hat{\bk}\) and \(\hat{\bm\xi}\), eliminating \(\delta\rho,\delta p,\delta\bB\) via Eqs.~\eqref{app:eq:cont}–\eqref{app:eq:deltaBxi}, and solving for \(\eta/\alpha\) gives

\begin{equation}\label{app:eq:beta_over_alpha_master}
\frac{\eta}{\alpha}
=\;\frac{k_\perp}{k_\parallel}\,
\frac{(c_{\mathrm s}^2+v_A^2)k^2-\omega'^2}{\,v_A^2 k^2-\omega'^2\,},
\end{equation}

where \(k_\parallel=\bk\!\cdot\!\hat{\bB}\) and \(k_\perp^2=k^2-k_\parallel^2\), \(v_A=|\bB|/\sqrt{\rho}\). Using the magnetosonic dispersion

\begin{equation}\label{app:eq:quartic_disp}
\omega'^4-(c_{\mathrm s}^2+v_A^2)k^2\,\omega'^2+
c_{\mathrm s}^2 v_A^2 k^2 k_\parallel^2=0 ,
\end{equation}
one recovers the compact identities quoted in Eq.~\eqref{eq:beta_alpha} after eliminating \(\omega'^2\) on the chosen branch, e.g.

\[
\frac{\eta}{\alpha}
= -\,\frac{\omega'^2 - c_{\mathrm s}^2 k^2 - v_A^2 k_\perp^2}{v_A^2 k_\parallel k_\perp}
= -\,\frac{v_A^2 k_\parallel k_\perp}{\omega'^2 - v_A^2 k_\parallel^2}\!,
\]

which are algebraically equivalent to \eqref{app:eq:beta_over_alpha_master} by the identity \( \omega'^2(\omega'^2-v_A^2 k^2)= c_{\mathrm s}^2 k^2(\omega'^2-v_A^2 k_\parallel^2) \) that follows from \eqref{app:eq:quartic_disp}. In the limits \(k_\perp\to 0\) or \(v_A\to 0\), \(\eta/\alpha\to 0\) as expected. In the quasi-perpendicular limit \(k_\parallel\to 0\), \(|\eta/\alpha|\to\infty\), to be read as \(\alpha\to 0\): the motion becomes purely \(\hat{\bm\xi}\)–polarized in the \((\bk,\bB)\) plane. The sign of \(\eta/\alpha\) is branch dependent (fast/slow) through the selected \(\omega'^2\).

\section{Driving Kernels and Obliquity Weights}\label{app:weights}

At fixed \((\omega,\bk_\perp)\), the coupled system
\begin{equation}\label{app:eq:block_system}
\mathsf{M}_{\mathrm{dd}}\,\mathbf{a}+\bm{m}_\zeta\,\zeta=\mathbf{f}_{\mathrm{d}},\qquad 
\bm{\ell}_{\mathrm{d}}^{\top}\mathbf{a}+\ell_\zeta\,\zeta=f_\zeta,
\end{equation}
implies \(\mathcal{Z}\) and \(\mathcal{S}\) as in Eq.~\eqref{eq:Z_S}. With upstream Alfv\'enic polarization \(\delta\bB_1^{(A)}=-\,\sigma\sqrt{\rho_1}\,\delta\bU_1^{(A)}\) (linear induction), each Alfv\'enic forcing term in \(\mathbf f_{\mathrm d}\) carries a factor \(\propto B_n\) from the boundary operators [cf. Eqs.~\eqref{eq:BC_tan_mom}–\eqref{eq:BC_tan_ind}]. After azimuthal averaging over the angle between \(\bk_\perp\) and \(\bB_{1\perp}\) (axisymmetric upstream spectra; Appendix~\ref{app:driver}), this factorization yields \(B_n^2=|\bB_1|^2\cos^2\theta_{Bn}\):
\begin{equation}\label{app:eq:SA_weight}
\big\langle|\mathcal S^{(A)}|^2\big\rangle=\mathcal G_A(\omega,\bk_\perp)\,E_A(\omega,\bk_\perp)\,\cos^2\theta_{Bn}.
\end{equation}
For compressive forcing, \(\delta p_1^{(C)}\) and \(\delta\bB_{1\perp}^{(C)}\) enter via tangential momentum/induction with magnitude set by \(|\bB_{1\perp}|\), and since \(|\bB_{1\perp}|^2=|\bB_1|^2\sin^2\theta_{Bn}\),
\begin{equation}\label{app:eq:SC_weight}
\big\langle|\mathcal S^{(C)}|^2\big\rangle
=\mathcal G_C(\omega,\bk_\perp)\,S_C(\omega,\bk_\perp)\,\sin^2\theta_{Bn}.
\end{equation}

Adding \eqref{app:eq:SA_weight}–\eqref{app:eq:SC_weight} gives Eq.~\eqref{eq:S_split}. Inserting into Eq.~\eqref{eq:zeta_general} and integrating over frequency along the transmitted fast branch yields the obliquity scaling in Eq.~\eqref{eq:obliquity_scaling}.

\section{Bordered Determinant, Grazing Linearization, and Response}\label{app:ZTmapping}

The impedance admits the bordered--determinant (Kreiss--Lopatinskii) form
\begin{equation}\label{app:eq:Z_bordered}
\mathcal{Z}(\omega,\bk_\perp)=
\frac{\det\!\begin{bmatrix}
\mathsf{M}_{\mathrm{dd}} & \bm{m}_\zeta\\
\bm{\ell}_{\mathrm{d}}^{\top} & \ell_\zeta
\end{bmatrix}}
{\det \mathsf{M}_{\mathrm{dd}}}\,,
\end{equation}
with downstream roots constrained by \(D_2(\omega-\bk\!\cdot\!\bU_2,\bk)=0\). Differentiating numerator and denominator with respect to the fast--branch normal root \(k_{n2}\) at fixed \((\omega,\bk_\perp)\), and using
\begin{equation}\label{app:eq:vg_identity}
\begin{aligned}
\frac{\partial_{k_{n2}} D_2}{\partial_{\omega'} D_2}
&= U_{n2}-v_{g,n2}(\omega,\bk_\perp)
= U_{n2}-\bn\!\cdot\!\frac{\partial \omega}{\partial \bk},\\[4pt]
\frac{\partial \omega}{\partial \bk}
&= \bU_2-\frac{\partial_{\bk} D_2}{\partial_{\omega'} D_2},
\end{aligned}
\end{equation}
relates variations along the fast normal root to variations of the normal group speed.

To make this connection explicit, introduce a coefficient \(C(\omega,\bk_\perp)\) that converts \(\partial_{k_{n2}}\) into \(\partial_{v_{g,n2}}\) by the chain rule implied by \eqref{app:eq:vg_identity}.

Let $k_{n2}=k_{n2}^{(g)}$ denote the downstream fast root at a grazing point, i.e.
$v_{g,n2}(\omega,\bk_\perp)=0$ on $D_2(\omega-\bk\!\cdot\!\bU_2,\bk)=0$.
A Taylor expansion in the fast normal root gives
\begin{equation}\label{app:eq:Z_kn_expand}
\mathcal{Z}(\omega,\bk_\perp)\;=\;\big(\partial_{k_{n2}}\mathcal{Z}\big)_{g}\,(k_{n2}-k_{n2}^{(g)})\;+\;\ii\,\Gamma\;+\;\cdots,
\end{equation}
and similarly
\begin{equation}\label{app:eq:vgn_kn_expand}
v_{g,n2}(\omega,\bk_\perp)\;=\;\big(\partial_{k_{n2}}v_{g,n2}\big)_{g}\,(k_{n2}-k_{n2}^{(g)})\;+\;\cdots.
\end{equation}

Eliminating $(k_{n2}-k_{n2}^{(g)})$ between \eqref{app:eq:Z_kn_expand} and \eqref{app:eq:vgn_kn_expand} yields

\begin{equation}\label{app:eq:Z_linear_correct}
\begin{aligned}
&\mathcal{Z}(\omega,\bk_\perp)\;=\;C(\omega,\bk_\perp)\,v_{g,n2}(\omega,\bk_\perp)\;+\;\ii\,\Gamma, \\
\qquad
&C\;\equiv\;\left.\frac{\partial_{k_{n2}}\mathcal{Z}}{\partial_{k_{n2}}v_{g,n2}}\right|_{g}.
\end{aligned}
\end{equation}

With $A\equiv \partial_{\omega'}D_2$ and $B\equiv \partial_{k_{n2}}D_2$, the normal group speed is
$v_{g,n2}=U_{n2}-B/A$.
Taking the total derivative along the dispersion surface and evaluating at grazing ($v_{g,n2}=0\Rightarrow B=U_{n2}A$)
gives the compact expression

\begin{equation}\label{app:eq:dvgn_dkn_grazing}
\begin{aligned}
&\left.\partial_{k_{n2}}v_{g,n2}\right|_{g}
= \\
&\ -\left.\frac{\partial_{k_{n2}k_{n2}}D_2
-2U_{n2}\,\partial_{k_{n2}\omega'}D_2
+U_{n2}^{2}\,\partial_{\omega'\omega'}D_2}{\partial_{\omega'}D_2}\right|_{g}.
\end{aligned}
\end{equation}

Compute $\partial_{k_{n2}}\mathcal{Z}$ from the $k_{n2}$--dependence of the fast-mode column in $\mathsf{M}_{\mathrm{dd}}$
(as in \eqref{app:eq:dZdk_rank1}), and divide by \eqref{app:eq:dvgn_dkn_grazing}:
\begin{equation}\label{app:eq:C_final}
C(\omega,\bk_\perp)
=\left.
\frac{\partial_{k_{n2}}\mathcal{Z}(\omega,\bk_\perp)}
{\partial_{k_{n2}}v_{g,n2}(\omega,\bk_\perp)}
\right|_{k_{n2}=k_{n2}^{(g)}}.
\end{equation}

An equivalent and directly computable representation follows by differentiating
\(\mathcal{Z}=\ell_\zeta-\bm{\ell}_{\mathrm d}^{\top}\mathsf{M}_{\mathrm{dd}}^{-1}\bm{m}_\zeta\) at fixed \((\omega,\bk_\perp)\).
The \(k_{n2}\)–dependence of \(\mathsf{M}_{\mathrm{dd}}\) enters through the downstream eigenmode columns; denote these by \(\bm c_j\) and let \(F\) be the transmitted fast--mode column. Since \(\bm{m}_\zeta\) and \(\bm{\ell}_{\mathrm d}\) do not depend on \(k_{n2}\), differentiating at fixed \((\omega,\bk_\perp)\) gives
\begin{equation}\label{app:eq:dZdk_pre}
\frac{\partial \mathcal{Z}}{\partial k_{n2}}
=\bm{\ell}_{\mathrm d}^{\top}\,\mathsf{M}_{\mathrm{dd}}^{-1}\,
\big(\partial_{k_{n2}}\mathsf{M}_{\mathrm{dd}}\big)\,
\mathsf{M}_{\mathrm{dd}}^{-1}\,\bm{m}_\zeta .
\end{equation}

If only the transmitted fast--mode column \(F\) depends on \(k_{n2}\) at the evaluation point, then \(\partial_{k_{n2}}\mathsf{M}_{\mathrm{dd}}=(\partial_{k_{n2}}\bm c_F)\,e_F^{\top}\), so
\begin{equation}\label{app:eq:dZdk_rank1}
\frac{\partial \mathcal{Z}}{\partial k_{n2}}
=\Big(\bm{\ell}_{\mathrm d}^{\top}\,\mathsf{M}_{\mathrm{dd}}^{-1}\,\partial_{k_{n2}}\bm c_F\Big)
\Big(e_F^{\top}\,\mathsf{M}_{\mathrm{dd}}^{-1}\,\bm{m}_\zeta\Big).
\end{equation}

Therefore the grazing linearization coefficient is
\begin{equation}\label{app:eq:C_rank1_correct}
C(\omega,\bk_\perp)
=\left.
\frac{
\Big(\bm{\ell}_{\mathrm d}^{\top}\,\mathsf{M}_{\mathrm{dd}}^{-1}\,\partial_{k_{n2}}\bm c_F\Big)
\Big(e_F^{\top}\,\mathsf{M}_{\mathrm{dd}}^{-1}\,\bm{m}_\zeta\Big)
}{
\partial_{k_{n2}}v_{g,n2}(\omega,\bk_\perp)
}
\right|_{k_{n2}=k_{n2}^{(g)}} ,
\end{equation}
with \(\partial_{k_{n2}}v_{g,n2}\) evaluated on the downstream fast branch (e.g., by the analytic expression in Eq.~\eqref{app:eq:dvgn_dkn_grazing} or by a centered difference taken along the fast normal root at fixed \((\omega,\bk_\perp)\)).
At grazing, Eq.~\eqref{app:eq:dvgn_dkn_grazing} may be inserted directly into the denominator of \eqref{app:eq:C_rank1_correct}.

The derivatives \(\partial_{k_{n2}}\bm c_F\) are taken at fixed \((\omega,\bk_\perp)\) along the downstream fast branch and can be formed analytically from the fast--mode polarization at the interface or numerically via a small centered difference in \(k_{n2}\). The vectors \(\mathsf{M}_{\mathrm{dd}}^{-1}\bm{m}_\zeta\) and \(\mathsf{M}_{\mathrm{dd}}^{-1}\partial_{k_{n2}}\bm c_F\) follow from two linear solves with the same \(\mathsf{M}_{\mathrm{dd}}\). This factorization is invariant under constant rescaling of the fast--mode column, provided the same normalization is used in \(\mathsf{M}_{\mathrm{dd}}\) and in its column derivative. If two transmitted modes approach grazing simultaneously, \(\partial_{k_{n2}}\mathsf{M}_{\mathrm{dd}}\) must be restricted to the subspace they span and \(e_F^{\top}\) replaced by the corresponding dual basis.

Therefore, to first order near grazing,
\begin{equation}\label{app:eq:Z_linear}
\mathcal{Z}(\omega,\bk_\perp)
= C(\omega,\bk_\perp)\,v_{g,n2}(\omega,\bk_\perp) + \ii\,\Gamma ,
\end{equation}

and the Lorentzian response

\begin{equation}\label{app:eq:T_Lorentz}
\begin{aligned}
&\ |\mathcal{T}(\omega,\bk_\perp)|^2
=\frac{1}{\,|C(\omega,\bk_\perp)|^2\,v_{g,n2}^2(\omega,\bk_\perp)+\Gamma^2\,},
\qquad \\
&\ \mathcal{T}=\frac{1}{\mathcal{Z}},
\end{aligned}
\end{equation}

follows. For dimensional consistency with \(C\,v_{g,n2}\), write \(\Gamma=|C|\,v_\nu\), where \(v_\nu\) is a physically motivated speed (e.g., along--front advection, leakage, or kinetic damping; Appendix~\ref{app:Gamma}). Then \( |\mathcal{T}|^2=\big(|C|^2[v_{g,n2}^2+v_\nu^2]\big)^{-1}\).

\section{Explicit Forms Used in the Schur Complement}\label{app:assembly}

For completeness, here we record the explicit rows of $\mathsf M_{\mathrm{dd}}$, the kinematic column $\bm m_\zeta$ (after $\partial_t\!\to -i\omega$, $\nabla_\perp\!\to i\bk_\perp$), and the upstream forcing vectors $(\mathbf f_{\mathrm d},f_\zeta)$ used in the Schur complement for $\mathcal Z$. 

% For the entropy/contact mode (denoted by $j=\mathrm{ent}$) we take the eigenvector at the interface to satisfy
% \begin{equation}\label{eq:entropy_mode_def}
% (\delta\bU_2)^{(\mathrm{ent})}=\boldsymbol{0},\qquad
% (\delta\bB_2)^{(\mathrm{ent})}=\boldsymbol{0},\qquad
% (\delta p_2)^{(\mathrm{ent})}=0,
% \end{equation}
% with $(\delta\rho_2)^{(\mathrm{ent})}\neq 0$ and $\delta S_2^{(\mathrm{ent})}$ chosen so that \eqref{eq:Lclosure} gives $(\delta p_2)^{(\mathrm{ent})}=0$.

Using \(\delta(\rho U_n)=U_n\,\delta\rho+\rho\,\delta U_n\) and the chosen normalization,

\begin{equation}\label{eq:M_mass}
(\mathsf{M}_{\mathrm{dd}})_{\mathrm{mass},j}
= U_{n2}\,(\delta\rho_2)^{(j)} + \rho_2\,(\delta U_{n2})^{(j)}.
\end{equation}

and from \(\bk\!\cdot\!\delta\bB=0\Rightarrow k_{n2}(\delta B_{n2})^{(j)}=-\bk_\perp\!\cdot\!(\delta\bB_{2\perp})^{(j)}\),

\begin{equation}\label{eq:M_Bn}
(\mathsf{M}_{\mathrm{dd}})_{B_n,j}=(\delta B_{n2})^{(j)}.
\end{equation}

For each admissible downstream mode $j$,
\begin{equation}\label{eq:M_energy}
(\bm{\ell}_{\mathrm d}^{\top})_j
= \big(\delta\bF_{E,n2}\big)^{(j)} \equiv \big(\delta\bF_{E,2}\cdot\bn\big)^{(j)} .
\end{equation}

The $\zeta$ coefficient and upstream forcing are

\begin{equation}\label{eq:ellzeta_fzeta_energy}
\begin{aligned}
&\ \ell_\zeta = i\omega\,\jump{Q_{E0}} - i\,\bk_\perp\!\cdot\!\jump{(\bF_{E0})_\perp},
\qquad \\
&\ f_\zeta = \delta\bF_{E,n1} \equiv \delta\bF_{E,1}\cdot\bn .   
\end{aligned}
\end{equation}

For tangential momentum, retaining both \(-(\delta B_n)\bB_\perp\) and \(-B_n\delta\bB_\perp\), and projecting along \(\btone,\bttwo\),

\begin{equation}\label{eq:M_tan_mom}
\begin{aligned}
(\mathsf{M}_{\mathrm{dd}})_{\mathrm{tm},j}^{(1)} 
&=
\btone\!\cdot\!\Big(
   \rho_2 U_{n2}\,(\delta\bU_{2\perp})^{(j)}
 + U_{n2}\,(\delta\rho_2)^{(j)}\,\bU_{2\perp} \\
& \quad + \rho_2\,(\delta U_{n2})^{(j)}\,\bU_{2\perp} \\
& \quad - (\delta B_{n2})^{(j)}\,\bB_{2\perp}
 - B_n\,(\delta\bB_{2\perp})^{(j)}
\Big),
\end{aligned}
\end{equation}

\begin{equation}\label{eq:M_tan_mom2}
\begin{aligned}
(\mathsf{M}_{\mathrm{dd}})_{\mathrm{tm},j}^{(2)} 
&=
\bttwo\!\cdot\!\Big(
   \rho_2 U_{n2}\,(\delta\bU_{2\perp})^{(j)}
 + U_{n2}\,(\delta\rho_2)^{(j)}\,\bU_{2\perp} \\
& \quad + \rho_2\,(\delta U_{n2})^{(j)}\,\bU_{2\perp} \\
& \quad - (\delta B_{n2})^{(j)}\,\bB_{2\perp}
 - B_n\,(\delta\bB_{2\perp})^{(j)}
\Big).
\end{aligned}
\end{equation}

For normal momentum, with \(\delta(\rho U_n^2)=U_n^2\delta\rho+2\rho U_n\delta U_n\) and \(\delta(B^2/2)=\bB\!\cdot\!\delta\bB\),

\begin{equation}\label{eq:M_norm_mom}
\begin{aligned}
(\mathsf{M}_{\mathrm{dd}})_{\mathrm{nm},j} 
&= U_{n2}^2\,(\delta\rho_2)^{(j)} + 2\rho_2 U_{n2}\,(\delta U_{n2})^{(j)} \\
&\quad + (\delta p_2)^{(j)}  + \bB_{2\perp}\!\cdot\!(\delta\bB_{2\perp})^{(j)}  - B_n\,(\delta B_{n2})^{(j)}.
\end{aligned}
\end{equation}

For tangential induction, using Eq.~\eqref{eq:RH_Etan},

\begin{equation}\label{eq:M_tan_ind}
\begin{aligned}
(\mathsf{M}_{\mathrm{dd}})_{\mathrm{ti},j}^{(1)} 
&= \btone\!\cdot\!\Big(
  (\delta U_{n2})^{(j)}\bB_{2\perp}
 + U_{n2}(\delta\bB_{2\perp})^{(j)} \\
&\quad - (\delta\bU_{2\perp})^{(j)}B_n
 - \bU_{2\perp}(\delta B_{n2})^{(j)}
\Big),
\end{aligned}
\end{equation}

\begin{equation}\label{eq:M_tan_ind2}
\begin{aligned}
(\mathsf{M}_{\mathrm{dd}})_{\mathrm{ti},j}^{(2)} 
&= 
\bttwo\!\cdot\!\Big(
   (\delta U_{n2})^{(j)}\bB_{2\perp}
 + U_{n2}(\delta\bB_{2\perp})^{(j)} \\
&\quad - (\delta\bU_{2\perp})^{(j)}B_n
 - \bU_{2\perp}(\delta B_{n2})^{(j)}
\Big).
\end{aligned}
\end{equation}

The \(\zeta\)-column follows by \(\partial_t\zeta\mapsto -\ii\omega\,\zeta\),

\begin{equation}\label{eq:m_zeta}
\bm{m}_\zeta=
\begin{bmatrix}
i\omega\,\jump{\rho}
- i\,\bk_\perp\!\cdot\!\jump{\rho_0\,\bU_{0\perp}}
\\[4pt]
-\,i\,\bk_\perp\!\cdot\!\jump{\bB_{0\perp}}
\\[4pt]
\displaystyle
i\omega\,\btone\!\cdot\!\jump{\rho_0\,\bU_{0\perp}}
- i\,\bk_\perp\!\cdot\!
\jump{\big(\bT_0^{\!\top}\btone\big)_{\!\perp}}
\\[4pt]
\displaystyle
i\omega\,\bttwo\!\cdot\!\jump{\rho_0\,\bU_{0\perp}}
- i\,\bk_\perp\!\cdot\!
\jump{\big(\bT_0^{\!\top}\bttwo\big)_{\!\perp}}
\\[4pt]
0
\\[4pt]
\displaystyle
i\omega\,\btone\!\cdot\!\jump{\bB_{0\perp}}
- i\,\bk_\perp\!\cdot\!
\jump{\big((\bU_0\bB_0-\bB_0\bU_0)^{\!\top}\btone\big)_{\!\perp}}
\\[4pt]
\displaystyle
i\omega\,\bttwo\!\cdot\!\jump{\bB_{0\perp}}
- i\,\bk_\perp\!\cdot\!
\jump{\big((\bU_0\bB_0-\bB_0\bU_0)^{\!\top}\bttwo\big)_{\!\perp}}
\end{bmatrix}
\end{equation}

The fifth component (normal momentum) is written as \(0\) because its kinematic contribution

\[
(\bm m_\zeta)_{\mathrm{nm}}
=
i\omega\,\bn\!\cdot\!\jump{\rho_0\,\bU_0}
- i\,\bk_\perp\!\cdot\!\jump{\big(\bT_0^{\!\top}\bn\big)_{\!\perp}}
\]

vanishes identically for a steady planar base shock: \(\jump{\rho_0 U_{n0}}=0\) and \(\jump{\big(\bT_0^{\!\top}\bn\big)_{\!\perp}}=0\) by the zeroth-order RH conditions.

The upstream forcing splits into Alfv\'enic and compressive contributions. For Alfv\'en (\(\delta\rho = 0 \)),

\begin{equation}\label{eq:fA_mass}
f^{(A)}_{\mathrm{mass}}=\,\rho_1\,\delta U_{n1}^{(A)},
\end{equation}

\begin{equation}\label{eq:fA_nm}
\begin{aligned}
f^{(A)}_{\mathrm{nm}}
&= 2\rho_1 U_{n1}\,\delta U_{n1}^{(A)}
+ \bB_{1\perp}\!\cdot\!\delta\bB_{1\perp}^{(A)}
- B_n\,\delta B_{n1}^{(A)} \\
&= (2\rho_1 U_{n1}+\sigma B_n\sqrt{\rho_1})\,\delta U_{n1}^{(A)} - \sigma\sqrt{\rho_1}\,\bB_{1\perp}\!\cdot\!\delta\bU_{1\perp}^{(A)}.
\end{aligned}
\end{equation}

\begin{equation}\label{eq:fA_tm}
\begin{aligned}
\boldsymbol{f}^{(A)}_{\mathrm{tm}}
&=\delta\!\big(\rho U_n \bU_\perp - B_n \bB_\perp\big)_1^{(A)} \\
&=
\rho_1 U_{n1}\,\delta\bU_{1\perp}^{(A)}
+\rho_1\,\delta U_{n1}^{(A)}\,\bU_{1\perp}
+U_{n1}\,\delta\rho_1^{(A)}\,\bU_{1\perp} \\
&\quad
- B_n\,\delta\bB_{1\perp}^{(A)}
-\delta B_{n1}^{(A)}\,\bB_{1\perp}.
\end{aligned}
\end{equation}

For Alfv\'en, $\delta\rho_1^{(A)}=0$ so the third term vanishes. Using \(\delta\bB_{1\perp}^{(A)}=-\sigma\sqrt{\rho_1}\,\delta\bU_{1\perp}^{(A)}\) we get,

\begin{equation}\label{eq:fA_tm_simplified}
\begin{aligned}
\boldsymbol{f}^{(A)}_{\mathrm{tm}}
&= \left(\rho_1 U_{n1}+\sigma B_n\sqrt{\rho_1}\right)\delta\bU_{1\perp}^{(A)} \\
&\quad +\rho_1\,\delta U_{n1}^{(A)}\,\bU_{1\perp}
-\delta B_{n1}^{(A)}\,\bB_{1\perp}.   
\end{aligned}
\end{equation}

\begin{equation}\label{eq:fA_ti}
\begin{aligned}
 \boldsymbol{f}^{(A)}_{\mathrm{ti}}
&=\delta\!\big(U_n\bB_\perp-\bU_\perp B_n\big)_1^{(A)} \\
&=
\delta U_{n1}^{(A)}\,\bB_{1\perp}
+U_{n1}\,\delta\bB_{1\perp}^{(A)} \\
&\quad -\delta\bU_{1\perp}^{(A)}\,B_n
-\bU_{1\perp}\,\delta B_{n1}^{(A)}.
\end{aligned}
\end{equation}

substitute \(\delta\bB_{1\perp}^{(A)}=-\sigma\sqrt{\rho_1}\,\delta\bU_{1\perp}^{(A)}\) to get

\begin{equation}\label{eq:fA_ti_simplified}
\begin{aligned}
\boldsymbol{f}^{(A)}_{\mathrm{ti}}
&= -\Big(\sigma U_{n1}\sqrt{\rho_1}+B_n\Big)\,\delta\bU_{1\perp}^{(A)} \\
&\quad +\delta U_{n1}^{(A)}\,\bB_{1\perp}
-\bU_{1\perp}\,\delta B_{n1}^{(A)}.
\end{aligned}
\end{equation}

For the compressive driver,

\begin{equation}\label{eq:fC_mass}
\begin{aligned}
f^{(C)}_{\mathrm{mass}}
&= 
   U_{n1}\,\delta\rho_1^{(C)} + \rho_1\,\delta U_{n1}^{(C)} \\
&= 
   \rho_1\,\delta U_{n1}^{(C)} + \rho_1 U_{n1}\frac{k}{\omega_1'}\,\alpha.
\end{aligned}
\end{equation}

\begin{equation}\label{eq:fC_nm}
\begin{aligned}
f^{(C)}_{\mathrm{nm}} 
&=\,U_{n1}^2\,\delta\rho_1^{(C)} + 2\rho_1 U_{n1}\,\delta U_{n1}^{(C)} + \delta p_1^{(C)} \\
& \quad + \bB_{1\perp}\!\cdot\!\delta\bB_{1\perp}^{(C)} - B_n\,\delta B_{n1}^{(C)}  \\
&= U_{n1}^2\,\frac{\rho_1 k}{\omega_1'}\,\alpha + 2\rho_1 U_{n1}\,\delta U_{n1}^{(C)} + c_{\mathrm s, 1}^2\,\frac{\rho_1 k}{\omega_1'}\,\alpha \\
& \quad  + \bB_{1\perp}\!\cdot\!\delta\bB_{1\perp}^{(C)} - B_n\,\delta B_{n1}^{(C)} .
\end{aligned}
\end{equation}

\begin{equation}\label{eq:fC_tm}
\begin{aligned}
\boldsymbol{f}^{(C)}_{\mathrm{tm}}
&=\delta\!\big(\rho U_n \bU_\perp - B_n \bB_\perp\big)_1^{(C)} \\
&=
\rho_1 U_{n1}\,\delta\bU_{1\perp}^{(C)}
+\rho_1\,\delta U_{n1}^{(C)}\,\bU_{1\perp}
+U_{n1}\,\delta\rho_1^{(C)}\,\bU_{1\perp} \\
&\quad
- B_n\,\delta\bB_{1\perp}^{(C)}
-\delta B_{n1}^{(C)}\,\bB_{1\perp}.
\end{aligned}
\end{equation}

\begin{equation}\label{eq:fC_ti}
\begin{aligned}
\boldsymbol{f}^{(C)}_{\mathrm{ti}}
&=\delta\!\big(U_n\bB_\perp-\bU_\perp B_n\big)_1^{(C)} \\
&=
\delta U_{n1}^{(C)}\,\bB_{1\perp}
+U_{n1}\,\delta\bB_{1\perp}^{(C)}
-\delta\bU_{1\perp}^{(C)}\,B_n
-\bU_{1\perp}\,\delta B_{n1}^{(C)}.
\end{aligned}
\end{equation}

If we work in the de Hoffmann–Teller / normal-incident frame with \(\bU_{1\perp}=0\), then the extra \(\delta U_n\) and \(\delta\rho\) terms vanish, but \(-\delta B_n\,\bB_{1\perp}\) generally does not unless we also have \(\bB_{1\perp}=0\).

\section{Normalization of Downstream Eigenmodes}\label{app:normalization}

For assembling $\mathsf M_{\mathrm{dd}}$ it is convenient to set $(\delta U_{n2})^{(j)}=1$ for each admissible mode $\psi_2^{(j)}$. A numerically robust alternative is to impose unit outgoing normal wave–action flux. On the dispersion surface $D_2(\omega',\bk)=0$, the wave–action density and flux scale as

\begin{equation}
\mathcal A_2^{(j)}\propto \partial_{\omega'}D_2\,\|\psi_2^{(j)}\|^2,
\qquad
\boldsymbol{\mathcal F}_2^{(j)}=\frac{\partial\omega}{\partial\bk}\,\mathcal A_2^{(j)}.
\end{equation}

Imposing unit outgoing normal flux gives

\begin{equation}\label{app:eq:unit_flux_norm}
\bn\!\cdot\!\boldsymbol{\mathcal F}_2^{(j)}=1
\ \implies \ 
\big|v_{g,n2}^{(j)}\,\partial_{\omega'}D_2\big|\,\|\psi_2^{(j)}\|^2=1,
\end{equation}

with the overall sign absorbed into the mode phase. Consequently,

\begin{equation}
\|\psi_2^{(j)}\|\;\sim\;\big(|v_{g,n2}^{(j)}|\,|\partial_{\omega'}D_2|\big)^{-1/2},
\end{equation}

so the amplitude grows like $|v_{g,n2}|^{-1/2}$ as $v_{g,n2}\!\to 0$. For numerical conditioning it is convenient to rescale each modal column by
\[
\sigma_j=\sqrt{|v_{g,n2}^{(j)}|\,|\partial_{\omega'}D_2|}\,,
\qquad
\tilde\psi_2^{(j)}=\sigma_j\,\psi_2^{(j)},
\]
which leaves the Schur complement and bordered determinant invariant but keeps column magnitudes $\mathcal O(1)$ near grazing. The grazing resonance therefore appears through $\mathcal Z$ (the bordered determinant) rather than through an arbitrary normalization factor, as standard in wave–action theory \cite[e.g.][]{BrethertonGarrett68,Whitham74}.

Regarding the invarinace of the bordered determinant, when evaluating $\partial_{k_{n2}}\bm c_F$ in the grazing linearization, the eigenmode normalization is held fixed with respect to $k_{n2}$. Equivalently, if a $k_{n2}$--dependent rescaling $\tilde{\bm c}_F=\sigma_F\,\bm c_F$ is used for numerical conditioning, then $\partial_{k_{n2}}\tilde{\bm c}_F$ is taken as $\sigma_F\,\partial_{k_{n2}}\bm c_F$ with $\sigma_F$ frozen at the evaluation point (no $\partial_{k_{n2}}\sigma_F$ term). This preserves the column-rescaling invariance of the linearized coefficient $C$.

\section{Effective Regularization \texorpdfstring{$\Gamma$}{Gamma}}\label{app:Gamma}

The small imaginary part in $\mathcal Z=C\,v_{g,n2}+\ii\,\Gamma$ is specified by a speed scale $v_\nu$,
\begin{equation}\label{app:eq:Gamma_dim}
\Gamma=|C(\omega,\bk_\perp)|\,v_\nu(\omega,\bk_\perp).
\end{equation}

We take $v_\nu=L_0\,\nu$ with a composite baseline damping rate

\begin{equation}\label{app:eq:vnu_from_rates}
\nu=\nu_{\rm tan\!:\!adv}+\nu_{\rm bulk~leak}+\nu_{\rm kin},
\end{equation}

which collects sinks only; particle or reflected–ion feedback that can reduce or reverse the net damping is incorporated in the main text via impedance correction terms (e.g., $\Lambda\,\mathcal T$), not added here.

Along–surface advection over an along–front coherence length $L_\parallel$ gives
\begin{equation}\label{app:eq:Gamma_tan}
\nu_{\rm tan\!:\!adv}\ \sim\ \frac{v_{g,\perp,2}}{L_\parallel}
\ \simeq\ \frac{c_{f2}\,\sin\theta_g}{L_\parallel}\,,
\end{equation}
where $\theta_g$ is defined by $v_{g,n2}=0$ and the last estimate is an order–of–magnitude fast–branch relation at grazing. Bulk leakage from sampling a finite normal–root width $\Delta k_{n2}$ yields the rate
\begin{equation}\label{app:eq:Gamma_bulk}
\nu_{\rm bulk\!:\!leak}\ \sim\
\bigg|\frac{\partial^2\omega}{\partial k_{n2}^2}\bigg|\,\Delta k_{n2}\,\frac{1}{\ell_n},
\end{equation}
where $\ell_n$ is an effective normal coherence length. Collisionless and non-ideal dissipation contribute

\begin{equation}\label{app:eq:Gamma_kin}
\begin{aligned}
\nu_{\rm kin}\ 
& \sim\
\nu_{\rm fast}(\beta,\theta_{kB},k\rho_s)\ \\
&\quad +\
\nu_{\rm slow}(\beta,\theta_{kB},k\rho_s)\ \\
&\quad +\
\nu_{\rm Hall/FLR}(k),
\end{aligned}
\end{equation}

with rates supplied by the chosen closure \cite[e.g.][]{melrose1986instabilities,Kulsrud2005}.

At $1$\,AU, reported local radii of curvature are $\sim3\times10^{6}$\,km (range $\sim3\times10^{5}$–$10^{7}$\,km) \cite{Neugebauer05}. Interpreting this as $L_\parallel$ and taking $v_{g,\perp,2}\sim200$–$400$\,km\,s$^{-1}$ gives the bracket
\[
\nu_{\rm tan\!:\!adv}\simeq \frac{v_{g,\perp,2}}{L_\parallel}\approx 1\times10^{-5}\text{ to }1\times10^{-3}\,\mathrm{s^{-1}}.
\]
In nondimensional variables with speed scaled by $U_{n1}$ and length by $L_0$,
\[
\Gamma=|C|\,\nu_\ast,\qquad \nu_\ast=\frac{L_0}{U_{n1}}\,\nu
\ \approx\ \frac{v_{g,\perp,2}}{U_{n1}}\frac{L_0}{L_\parallel}.
\]
For $L_\parallel\sim 3\times 10^{5}\text{--}10^{7}\,$km, $v_{g,\perp,2}\sim 200\text{--}400$\,km\,s$^{-1}$, and $L_0\sim10^{5}\text{--}10^{6}\,$km with $v_{g,\perp,2}/U_{n1}\sim\mathcal O(1)$, one obtains $\nu_\ast\sim 5\times10^{-3}$ up to $\mathcal O(1)$ (and can exceed unity for the smallest $L_\parallel$ and largest $L_0$). In applications where the grazing resonance is treated as a narrow feature, we restrict to $\nu_\ast\ll 1$ so that $\Gamma$ acts as a weak regularization rather than dominating the response.

The effective regularization used in Eq.~\eqref{app:eq:Z_linear} is therefore

\begin{equation}\label{app:eq:Gamma_final}
\Gamma(\omega,\bk_\perp)=|C(\omega,\bk_\perp)|\,
\Big[L_0\big(\nu_{\rm tan\!:\!adv}+\nu_{\rm bulk\!:\!leak}+\nu_{\rm kin}\big)\Big],
\end{equation}

where $\Gamma$ here encodes baseline damping; any additional particle/reflection feedback enters through the impedance correction in the main text. $\Gamma$ sets the angular/temporal breadth of the near–grazing selection and thus the width of the Lorentzian response in Eq.~\eqref{eq:T_Lorentz}.

\section{Injected Turbulence and Polarization Content}\label{app:driver}

The upstream driver is a statistically stationary, mean-zero field convected by $\bU_1$ with $\omega=\bk\!\cdot\!\bU_1$. One-dimensional perpendicular spectra over two decades are taken as

\begin{equation}\label{app:eq:1D_perp_spectra}
\begin{aligned}
&\ E_A(k_\perp)=C_A\,k_\perp^{-5/3},\qquad \\
&\ S_C(k_\perp)=C_C\,k_\perp^{-5/3},\qquad 
k_{\perp,\min}\le k_\perp\le k_{\perp,\max}, 
\end{aligned}
\end{equation}

normalized to achieve target rms amplitudes and the compressive fraction $\chi_C=S_C^{\rm tot}/(E_A^{\rm tot}+S_C^{\rm tot})$.

Power in $(k_\perp,k_\parallel)$ is distributed with respect to the upstream field $\bB_1$ by a critically balanced envelope with lognormal scatter,

\begin{equation}\label{app:eq:2D_envelope}
\mathcal{G}(k_\parallel;k_\perp)
=\frac{1}{\sqrt{2\pi}\,\sigma}\,
\exp\!\left[-\frac{\big(\ln[k_\parallel/k_{\rm cb}(k_\perp)]\big)^2}{2\sigma^2}\right]\,
\frac{1}{k_\parallel},
\end{equation}

\[k_{\rm cb}(k_\perp)=C_{\rm cb}\,k_\perp^{2/3},\]

To define spectra on $k_\parallel\in(-\infty,\infty)$ we use the even extension

\begin{equation}
\begin{aligned}
& \mathcal{G}_{\rm even}(k_\parallel;k_\perp)
=\frac12\,\mathcal{G}(|k_\parallel|;k_\perp), \\
& \int_{-\infty}^{\infty}\mathcal{G}_{\rm even}(k_\parallel;k_\perp)\,dk_\parallel=1.   
\end{aligned}
\end{equation}

so that $\int_0^\infty \mathcal{G}(k_\parallel;k_\perp)\,dk_\parallel=1$. This yields the 3-D PSDs

\begin{equation}\label{app:eq:3D_PSDs}
\begin{aligned}
\mathcal{P}_A(\bk)&=\frac{E_A(k_\perp)}{2\pi k_\perp}\,\mathcal{G}_{\rm even}(k_\parallel;k_\perp),\\
\mathcal{P}_C(\bk)&=\frac{S_C(k_\perp)}{2\pi k_\perp}\,\mathcal{G}_{\rm even}(k_\parallel;k_\perp).
\end{aligned}
\end{equation}

which satisfy $\int_0^{2\pi}\!d\phi\int_{-\infty}^{\infty}\!dk_\parallel\,\mathcal{P}_{A,C}\,k_\perp=E_{A,C}(k_\perp)$.

Alfv\'enic polarization is $\delta\bU_1^{(A)}(\bk)=\delta U_1^{(A)}(\bk)\,\hat{\bm e}_A$ with $\hat{\bm e}_A=(\bk\times\bB_1)/|\bk\times\bB_1|$ and
\begin{equation}\label{app:eq:Alf_sign}
\delta\bB_1^{(A)}=-\,\sigma\,\sqrt{\rho_1}\,\delta\bU_1^{(A)},
\end{equation}
consistent with the linear induction relation (with $\sigma=\pm1$ the Alfv\'en-branch sign). The set where $\bk\times\bB_1=\boldsymbol{0}$ has zero measure in the axisymmetric construction. Compressive polarization follows Appendix~\ref{app:pol}, with $\delta\bU_1^{(C)}=\alpha\,\hat{\bk}+\eta\,\hat{\bm\xi}$ and $\eta/\alpha$ given by Eq.~\eqref{app:eq:beta_over_alpha_master}; the linear relations \eqref{eq:Lmass}–\eqref{eq:Lclosure} then determine $\delta\rho_1^{(C)}$, $\delta p_1^{(C)}$, and $\delta\bB_1^{(C)}$ used in the forcing vectors.

Since $E_A,S_C\propto k_\perp^{-5/3}$ while the along--front advective rate scales as $\nu_{\rm tan\!:\!adv}\sim v_{g,\perp,2}/L_\parallel$ (Appendix~\ref{app:Gamma}), taking $L_\parallel\propto 1/k_\perp$ (coherence set by the along--front wavelength of each mode) gives $\nu_{\rm tan\!:\!adv}\propto k_\perp$, so the largest accessible perpendicular scales carry the dominant corrugation energy---consistent with the morphology of the synthetic $\zeta$ maps.

\section{Directional Measure and Residence Time at Grazing}\label{app:geom_residence}

For the downstream fast branch with $\omega=\omega'_2+\bk\!\cdot\!\bU_2$ and $\omega'_2\simeq \pm c_{f2}\,k$, the group velocity is $\bv_{g,2}\simeq \pm c_{f2}\,\hat{\bk}+\bU_2$. Let $\theta$ be the angle between $\hat{\bk}$ and $\bn$ and $\theta_g$ the solution of

\begin{equation}\label{app:eq:def_thetag}
\begin{aligned}
v_{g,n2}(\theta_g)=\bn\!\cdot\!\bv_{g,2} \;=\; U_{n2} \pm c_{f2}\cos\theta_g \;=\; 0 \\
\ \Rightarrow\
\cos\theta_g=\mp\,\frac{U_{n2}}{c_{f2}},\qquad
\sin\theta_g=\sqrt{1-\frac{U_{n2}^2}{c_{f2}^2}}.  
\end{aligned}
\end{equation}

At fixed $k$ and for an isotropic distribution of directions, the purely geometric directional measure of the grazing set follows from

\begin{equation}\label{app:eq:geom_factor}
\begin{aligned}
&\ \int \delta\!\big(v_{g,n2}(\theta)\big)\,d\Omega = \\
&\ \int_0^{2\pi}d\phi\int_0^\pi \sin (\theta) d\theta \, \delta \left(U_{n2} \pm c_{f2}\cos\theta \right) = \frac{2\pi}{c_{f2}}.   
\end{aligned}
\end{equation}

Thus, at fixed $k$, the geometric factor contributes $\propto c_{f2}^{-1}$, consistent with the discussion following Eq.~\eqref{eq:kperp_int}.

Complementarily, the residence-time weighting for a packet of normal extent $\ell_n$ convected by $U_{n2}$ near the interface is
\begin{equation}\label{app:eq:dwell}
\tau_{\rm dwell}\sim \frac{\ell_n}{|U_{n2}|},\qquad
\mathcal{R}\ \equiv\ \frac{v_{g,\perp,2}}{|U_{n2}|}\ \simeq\ \frac{c_{f2}\,\sin\theta_g}{|U_{n2}|}.
\end{equation}
When statistics weight along–surface exposure, $\mathcal R$ multiplies the angular factor; in the limit $|U_{n2}|\ll c_{f2}$ one has $\sin\theta_g\to 1$ and $\mathcal R\propto c_{f2}/|U_{n2}|$, consistent with Eq.~\eqref{eq:kperp_int}.

The angular $\delta$–measure above should be understood as the narrow–resonance limit; in practice, the finite regularization $\Gamma=|C|\,v_\nu$ (Appendix~\ref{app:Gamma}) broadens this selection in $\theta$ with local width $\Delta\theta\sim v_\nu/(c_{f2}\,\sin\theta_g)$, as captured by the Lorentzian response in Eq.~\eqref{eq:T_Lorentz}.

\section{Reaction-Diffusion Model, Planar DSA Normalization, and \texorpdfstring{$\mathcal{T}_{\rm cr}$}{Tcr}}
\label{app:Tcr_derivation}

We collect the reaction–diffusion closure that leads to Eq.~\eqref{eq:accel_Tcr} and its normalization by the planar diffusive–shock–acceleration (DSA) time. The energetic (cosmic–ray) pressure (or normal–stress) perturbation on the surface, denoted $\delta P_{\rm cr}(\boldsymbol{x}_\perp,t)$, is modeled as a linear, along–surface reaction–diffusion field driven by corrugation–induced injection modulations:
\begin{equation}
\label{app:eq:cr_react_diffusion}
\partial_t \delta P_{\rm cr}
= -\,\frac{1}{t_{\rm acc}}\,\delta P_{\rm cr}
\;+\;\mathcal K\,\nabla_{\!\perp}^2 \delta P_{\rm cr}
\;+\;\mathcal A_{\rm cr}\,\delta\Psi,
\end{equation}
where $\mathcal K$ is an effective tangential (along-front) diffusivity, $t_{\rm acc}$ is a single acceleration/relaxation time, and $\mathcal A_{\rm cr}$ sets the normalization from injected fraction to normal stress.

With the plane–wave convention $\exp\{\ii(\bk\!\cdot\!\boldsymbol{x}-\omega t)\}$, one has $\partial_t\mapsto -\ii\omega$ and $\nabla_{\!\perp}^2\mapsto -k_\perp^2$, so the Fourier transform of \eqref{app:eq:cr_react_diffusion} gives
\begin{equation}
\label{app:eq:cr_fourier}
\big(-\ii\omega + t_{\rm acc}^{-1} + \mathcal K k_\perp^2\big)\,\delta P_{\rm cr}(\omega,\bk_\perp)
= \mathcal A_{\rm cr}\,\delta\Psi(\omega,\bk_\perp).
\end{equation}

The source $\delta\Psi$ is set by the corrugation–induced obliquity modulation from Eqs.~\eqref{eq:accel_dtheta}–\eqref{eq:accel_deltaeta}. Using
\begin{equation}
\label{app:eq:deltaeta_repeat}
\delta\Psi_{\bk_\perp}
= \ii\,\Psi'(\theta_{Bn})\,
\frac{\bB_{1\perp}\!\cdot\!\bk_\perp}{|\bB_1|\,\sin\theta_{Bn}}\,
\zeta_{\bk_\perp},
\end{equation}
we collect constants into
\begin{equation}
\label{app:eq:Lambda_cr_def}
\Lambda_{\rm cr}(\bk_\perp)
= \ii\,\mathcal A_{\rm cr}\,\Psi'(\theta_{Bn})\,
\frac{\bB_{1\perp}\!\cdot\!\bk_\perp}{|\bB_1|\,\sin\theta_{Bn}},
\end{equation}
so that $\mathcal A_{\rm cr}\,\delta\Psi=\Lambda_{\rm cr}\,\zeta$. Solving \eqref{app:eq:cr_fourier} for $\delta P_{\rm cr}$ and factoring $t_{\rm acc}$ yields
\begin{equation}
\label{app:eq:Tcr_from_PDE}
\delta P_{\rm cr}(\omega,\bk_\perp)
= \frac{t_{\rm acc}}{\,1-\ii\omega t_{\rm acc}+\mathcal K t_{\rm acc} k_\perp^2\,}\;
\Lambda_{\rm cr}(\bk_\perp)\,\zeta(\omega,\bk_\perp),
\end{equation}
which identifies the transfer function (Eq.~\eqref{eq:accel_Tcr}),
\begin{equation}
\label{app:eq:Tcr_box}
\mathcal{T}_{\rm cr}(\omega,\bk_\perp)
= \frac{t_{\rm acc}}{\,1-\ii\omega t_{\rm acc}+\mathcal K t_{\rm acc} k_\perp^2\,}.
\end{equation}

The remaining ingredient is $t_{\rm acc}$. In the test–particle, planar DSA limit the acceleration time is given by \cite{Drury83}

\begin{equation}
\label{app:eq:tacc_general}
t_{\rm acc}
=\frac{3}{U_{n1}-U_{n2}}
\left(\frac{\kappa_1}{U_{n1}}+\frac{\kappa_2}{U_{n2}}\right),
\end{equation}

with $U_{n1}$ and $U_{n2}$ the upstream and downstream normal speeds in the shock frame and $\kappa_{1,2}$ the corresponding diffusion coefficients. Using $U_{n2}=U_{n1}/r$ and, for simplicity, $\kappa_1=\kappa_2=\kappa$,
\begin{equation}
\label{app:eq:tacc_equal_kappa}
\begin{aligned}
t_{\rm acc}
&=\frac{3}{U_{n1}-U_{n1}/r}\left(\frac{\kappa}{U_{n1}}+\frac{\kappa}{U_{n1}/r}\right) \\
&=\frac{3}{U_{n1}(1-1/r)}\left(\frac{\kappa}{U_{n1}}+\frac{\kappa r}{U_{n1}}\right)\\
&=\frac{3}{U_{n1}}\frac{r}{r-1}\,\frac{\kappa(r+1)}{U_{n1}}
=\frac{\kappa}{U_{n1}^2}\,\frac{3r(r+1)}{r-1} \\
&\equiv\;\chi_{\rm acc}\,\frac{\kappa}{U_{n1}^2},
\end{aligned}
\end{equation}
with $\chi_{\rm acc}=3r(r+1)/(r-1)$. If $\kappa_1\neq\kappa_2$, \eqref{app:eq:tacc_general} should be retained; the subsequent algebra is unchanged.%
% \footnote{If diffusion is anisotropic, replace $\kappa$ by the appropriate along–front projection of $(\kappa_\parallel,\kappa_\perp)$; this modifies only the numerical value of $t_{\rm acc}$ and the $k_\perp$–filter $1+\kappa t_{\rm acc}k_\perp^2$ in \eqref{app:eq:Tcr_box}.}

To connect with the scalar interface equation, recall $\mathcal{Z}\,\zeta=\mathcal S$ from Eq.~\eqref{eq:Z_S}. The energetic component contributes an additional compressive drive $S_{\rm cr}=\Lambda_{\rm cr}\,\mathcal{T}_{\rm cr}\,\zeta$ on the right–hand side. Moving this term to the left renormalizes the impedance,
\begin{equation}
\label{app:eq:Zeff_appendix}
\big(\mathcal{Z}-\Lambda_{\rm cr}\,\mathcal{T}_{\rm cr}\big)\,\zeta
=\mathcal S,
\end{equation}
and, near grazing where $\mathcal Z=C\,v_{g,n2}+\ii\Gamma$ [Eq.~\eqref{eq:Z_linear}], this reproduces Eq.~\eqref{eq:accel_Zeff}. Evaluating $\mathcal{T}_{\rm cr}$ on the drift of corrugations $\omega\simeq k_\perp v_{\rm corug}$ then gives the along–front hot–spot scaling quoted in Eq.~\eqref{eq:accel_kstar}.

\bibliography{apssamp}% Produces the bibliography via BibTeX.

\end{document}